%% file: main.tex
\documentclass[letterpaper,twocolumn,10pt,anonymous]{article}
\usepackage{usenix}
\usepackage{savesym}
\usepackage[T1]{fontenc}
\restoresymbol{TXF}{iint}
\usepackage{amsfonts}
\usepackage{amsmath}
\usepackage{textcomp}
\usepackage{bbding}
\usepackage{xspace}
\usepackage{pifont}
\usepackage[normalem]{ulem}
\usepackage{hyphenat}
\usepackage{algorithm}
\usepackage{algorithmicx}
\usepackage[noend]{algpseudocode}
\usepackage{mathtools}
\usepackage{xspace}
\usepackage{enumitem}
\usepackage{titlesec}
\usepackage{multirow}
\usepackage{todonotes}
\usepackage{balance}
\usepackage{booktabs}
\usepackage{titlesec}

\titlespacing*{\section}{0pt}{4pt plus 2pt minus 1pt}{4pt plus 2pt minus 1pt}
\titlespacing*{\subsection}{0pt}{2pt plus 2pt minus 0pt}{2pt plus 2pt minus 0pt}
\setlist{nosep}
\newcommand{\rpt}{{\sf report}\xspace}
\newcommand{\pro}{{\sf c-propose}\xspace}
\newcommand{\prp}{{\sf c-prepare}\xspace}
\newcommand{\cmt}{{\sf c-commit}\xspace}
\newcommand{\cpt}{{\sf c-checkpoint}\xspace}
\newcommand{\vc}{{\sf c-view-change}\xspace}
\newcommand{\nv}{{\sf c-new-view}\xspace}
\newcommand{\init}{{\sf init history}\xspace}
\newcommand{\req}{{\sf request}\xspace}
\newcommand{\one}{{\sf pre-prepare}\xspace}
\newcommand{\two}{{\sf prepare}\xspace}
\newcommand{\three}{{\sf commit}\xspace}
\newcommand{\reply}{{\sf reply}\xspace}
\newcommand{\order}{{\sf order}\xspace}
\newcommand{\share}{{\sf sign-share}\xspace}

\newcommand{\PRO}{\footnotesize \textsf{C-PROPOSE}\xspace}

\newcommand{\abs}{{\sf Abstract}\xspace}
\newcommand{\adapt}{{\sc Adapt}\xspace}
\newcommand{\adaptsharp}{{\sc Adapt}\#\xspace}

\colorlet{boxcolor}{teal!10!white}

\newtheorem{metalemma}{Lemma}[section]

\newtheorem{Lemma}[metalemma]{Lemma}
\newtheorem{Theorem}[metalemma]{Theorem}

\newenvironment{proof}{\noindent{\bf Proof:}\rm}

\def\sys{BFTBrain\xspace}

\newcommand{\sparagraph}[1]{\vspace{1mm}\noindent {\bf #1}\xspace}

\DeclareMathOperator*{\argmax}{arg\,max}

\newif\ifremove
\removefalse

\begin{document}
\title{\Large \bf \sys: Adaptive BFT Consensus with Reinforcement Learning}

\author{
{\rm Chenyuan Wu} \\ University of Pennsylvania
\and {\rm Haoyun Qin} \\ University of Pennsylvania 
\and {\rm Mohammad Javad Amiri} \\ Stony Brook University
\and {\rm Boon Thau Loo} \\ University of Pennsylvania
\and {\rm Dahlia Malkhi} \\ UC Santa Barbara
\and {\rm Ryan Marcus} \\ University of Pennsylvania
}

\maketitle
\input{sec_abstract}
\input{sec_intro}

\input{sec_landscape}
\input{sec_overview}

\input{sec_learning}

\input{sec_collect}
\input{sec_impl}
\input{sec_eval}

\input{sec_related}
\input{sec_conc}


\balance
\bibliographystyle{plain}
\bibliography{_blockchain,_privacy,_system,Ryan-cites-long,_learning}

\clearpage
\nobalance
\appendix
\input{seca_background}

\input{seca_switch}

\input{seca_collect}

\input{seca_results}

\end{document}

%% file: sec_abstract.tex
\begin{abstract}
This paper presents \sys, a reinforcement learning (RL) based Byzantine fault-tolerant (BFT) system that provides significant operational benefits: a plug-and-play system suitable for a broad set of hardware and network configurations, and adjusts effectively in real-time to changing fault scenarios and workloads. 
\sys adapts to system conditions and application needs by switching between a set of BFT protocols in real-time. Two main advances contribute to \sys's agility and performance. First, \sys is based on a systematic, thorough modeling of metrics that correlate the performance of the studied BFT protocols with varying fault scenarios and workloads. These metrics are fed as features to \sys's RL engine in order to choose the best-performing BFT protocols in real-time. Second, \sys coordinates RL in a decentralized manner which is resilient to adversarial data pollution, where nodes share local metering values and reach the same learning output by consensus. 
As a result, in addition to providing significant operational benefits, \sys improves throughput over fixed protocols by $18\%$ to $119\%$ under dynamic conditions and outperforms state-of-the-art learning based approaches by $44\%$ to $154\%$.



\end{abstract}

%% file: sec_intro.tex
\section{Introduction} \label{sec:intro}

Byzantine fault-tolerant (BFT) consensus protocols are the core engines powering the state machine replication (SMR) paradigm, ensuring that non-faulty replicas execute client requests in the same order, despite the existence of $f$ Byzantine replicas. The ability to tolerate up to $f$ arbitrary failures makes BFT protocols a key component in various distributed systems, including permissioned blockchains~\cite{kwon2014tendermint, androulaki2018hyperledger}, distributed file systems~\cite{castro2002practical, clement2009upright}, locking services~\cite{clement2009making}, firewalls~\cite{garcia2013intrusion, garcia2016sieveq}, key-value stores~\cite{goodson2004efficient, dobre2013powerstore}, and SCADA systems~\cite{babay2019deploying, kirsch2013survivable}.  

While various BFT protocols have been proposed (e.g., PBFT \cite{castro1999practical, castro2002practical}, Zyzzyva \cite{kotla2007zyzzyva}, CheapBFT \cite{kapitza2012cheapbft}, Prime \cite{amir2011prime}, SBFT \cite{gueta2019sbft} and HotStuff-2 \cite{malkhi2023hotstuff}), there is no one-size-fits-all solution. The performance ranking of BFT protocols varies significantly depending on client workloads, network configurations, and application needs. 
For example, protocols that reduce message complexity by increasing communication phases exhibit better throughput but worse latency. In addition, adversarial behaviors in the system also affect the best-performing protocol choice. The lack of a clear ``winner'' among BFT protocols makes it difficult for application developers to choose one and may invalidate their choice if workloads or attacks change. This is exacerbated in blockchain systems where application workloads and potential attacks are diverse and dynamic~\cite{gramoli2023diablo}.

To address this challenge, two prior systems, \abs \cite{guerraoui2010next,aublin2015next} and \adapt \cite{bahsoun2015making} have been proposed. Both systems combine multiple protocols under a single framework and switch between protocols adaptively in real-time. Specifically, \abs creates an adaptive framework whereby there is a predefined switching order of BFT protocols. If one protocol does not make sufficient progress, it is aborted and the next protocol in the predefined order is selected. Such an approach lacks flexibility and is unlikely to work in all scenarios. 

\adapt enhances protocol selection using supervised learning, but it faces operational limitations. First, it relies on a single replica to collect data, train the machine learning model, and then distribute the decision to all other replicas. In a Byzantine environment, such centralized control is not realistic and runs counter to the entire BFT approach. Second, its requirement for prolonged upfront data collection and supervised learning hinders its ability to adapt to unforeseen conditions and workloads. Lastly, as our experiments will demonstrate, \adapt's incomplete feature space design renders it ineffective under diverse fault scenarios.

To embrace the idea of a multi-protocol BFT engine while considerably enhancing its practicality, we propose {\em \sys}, a reinforcement learning (RL) based BFT system. At a high level, given a performance metric to optimize, \sys smartly switches between a set of BFT protocols at run-time under dynamic workloads. \sys is \textit{practical} in two aspects --- it not only maintains adaptivity under various adversarial behaviors, but also provides significant \textit{operational benefits} when deployed on different hardware and system configurations: instead of manually choosing from multiple alternative systems for deployment, or running a prolonged data collection process prior to the deployment, \sys only requires running one system that automatically re-configures itself to implement a top-performing protocol in real-time.

To leverage RL, \sys measures in real-time performance metrics obtained locally by replicas. Beyond standard metrics like commit throughput, \sys employs fine-grained metrics that offer deeper performance insights, such as the ratio of requests that are committed in the fast path of dual-path protocols, the number of received messages per slot, and the interval between consecutive leader proposals. These new metrics are measured in a distributed manner and serve as features for \sys's RL engine. By modeling the selection of a BFT protocol as a contextual multi-armed bandit problem, the RL engine strategically tests different protocols at run-time to learn which ones are well-suited to the current system conditions. \sys coordinates RL in a decentralized manner, where nodes share local features/rewards by consensus and reach the same learning output, achieving resilience to adversarial data pollution. Our extensive evaluation shows that \sys significantly outperforms fixed protocols and existing adaptive systems in dynamic workloads and faults.

Specifically, this paper makes the following contributions.

\noindent {\bf BFT based on reinforcement learning.} To our best knowledge, \sys is the first BFT system that utilizes reinforcement learning to achieve run-time adaptability. It does not rely on a lengthy data collection process prior to every deployment and effectively adapts to unforeseen system conditions.

\noindent {\bf Adaptivity to fault scenarios.} \sys is the first BFT consensus system that not only automatically adapts to dynamic user workloads under a broad set of hardware and system configurations, but also adapts to various fault scenarios. 

\noindent {\bf Dissecting BFT performance.} Through a suite of workload and fault parameters, we conducted a large-scale study examining the relationship between protocol algorithmic design and performance, encompassing a wide range of well-known BFT protocols, including PBFT, Zyzzyva, CheapBFT, Prime, SBFT, and HotStuff-2. Our experiments highlight the vast state space, rendering existing solutions impractical.
    
\noindent {\bf Prototype and experimental evaluation.} We developed a prototype of \sys and integrated it with the six BFT protocols listed above. Our CloudLab evaluation demonstrates that \sys improves throughput by $18\%$-$119\%$ over fixed protocols under dynamic conditions and outperforms the state-of-the-art learning-based approach by $44\%$-$154\%$ depending on how many data are polluted by adversaries.





%% file: sec_landscape.tex
\section{Landscape of BFT Performance} \label{sec:landscape}

To motivate \sys, we first conduct an in-depth performance study to examine how various BFT protocols perform under different conditions. The key takeaway is that no single BFT protocol is always ``better'' or ``worse'' than others, but rather that different protocols perform better/worse than others under different circumstances. 

\subsection{Comparing Representative BFT Protocols}

We picked six representative BFT protocols as our exploration targets: PBFT~\cite{castro1999practical}, Zyzzyva~\cite{kotla2007zyzzyva}, CheapBFT~\cite{kapitza2012cheapbft}, Prime~\cite{amir2011prime}, SBFT~\cite{gueta2019sbft}, and HotStuff-2~\cite{malkhi2023hotstuff}. All six protocols are \textit{leader-based}, working in the partial synchrony settings with networks of  $n=3f+1$  nodes\footnote{CheapBFT requires only $2f+1$ with the help of trusted hardware, but was evaluated with $3f+1$ for simplicity. Correctness is still guaranteed.}. 
Detailed background on these protocols is provided in Appendix~\ref{sec:appendix_background}. These protocols and the benchmarking tools were implemented under a common software framework Bedrock~\cite{amiri2024bedrock}; hence, our evaluation below focuses on the impact of algorithmic logic of these protocols on performance, rather than the effect of the implementation details of a specific system. 

Our experiments were conducted on CloudLab~\cite{duplyakin2019design} where each replica is an \texttt{xl170} bare metal server. We ran these six protocols under diverse workloads and fault scenarios, and compared their average throughput during $120$ second runs that were stably reproduced $10$ times. For a fair comparison, the common internal parameters (e.g., batch size and view-change timer) of all six protocols were configured with the same values. For simplicity and facilitating protocol switching, as described in later sections, we added $f$ extra replicas to the original CheapBFT acting as \emph{active} replicas (see~\cite{kapitza2012cheapbft}). This approach affects the hardware resource consumption and scalability of the original CheapBFT, but does not change its performance under our setups. We also emulated the overhead of the trusted subsystem CASH by injecting $60$ $\mu s$ delay for creating and verifying message certificates. 

\begin{table}[t]
\caption{Our observed best-performing BFT protocols under different conditions. The advantage over the second-best protocol is shown in the last column.}
\vspace{1em}
\scriptsize
\begin{tabular}{ccccc}
\toprule
$f$ & \# of absentees & request size & proposal slowness  & best protocol \\ \midrule

1  & 0  &4KB & 0ms & Zyzzyva (15.6\%)  \\ 
4  & 0  &4KB & 0ms & Zyzzyva (34.3\%) \\
4  & 0  &100KB & 0ms & CheapBFT (8.5\%)\\ 
4  & 4  &4KB & 0ms & CheapBFT (13.1\%)  \\ 
4  & 0  &0KB & 20ms & HotStuff-2 (45.4\%)\\
4  & 0  &1KB & 20ms & HotStuff-2 (44.8\%)\\ 
4  & 0  &0KB & 100ms & Prime (16.9\%)\\ 
1  & 0  &0KB & 20ms & Prime (71.5\%) \\ 
\bottomrule
\end{tabular}
\vspace{1em}
\label{tbl:landscape}
\end{table}

Table~\ref{tbl:landscape} summarizes our evaluation results. Importantly, it demonstrates that \textbf{no single protocol dominates in all conditions}. The first four columns indicate the workloads and fault scenarios: system size, number of non-responsive nodes, client request size, and slowness of leader proposals; a more detailed description is provided in Section~\ref{sec:state_action}. The last column shows the best-performing protocol under each condition as well as its relative percentage advantage in throughput over the second-best protocol; the throughput of all six protocols under each condition is provided in Appendix~\ref{sec:appendix_table}.

It is worth underscoring that a ``slowness-attack'', capturing the interval between two consecutive leader proposals, appears in many studies~\cite{clement2009making, aublin2015next, amir2011prime}. It captures a Byzantine attack where a malicious leader deliberately postpones its proposal before the view-change timer expires. In addition, it could also happen naturally in a heterogeneous deployment, where a leader has weaker compute or network resources compared to other replicas. 
Below, we navigate through the table and explain the insights behind the results. 

\noindent {\bf Row 1-3.} The first three rows are in an ideal world where every replica is benign and responsive, with no obvious slowness in leader proposals, while varying the network and request sizes. When $n=4$ ($f=1$) and the request size is $4$KB, Zyzzyva outperforms the next best protocol CheapBFT by $15.6\%$. When increasing the network size to $n=13$ ($f=4$), Zyzzyva outperforms CheapBFT by $34.3\%$. However, when the request size increases to 100KB, a flip of ranking occurs: CheapBFT becomes the best-performing protocol and outperforms the second best protocol HotStuff-2 by $8.5\%$, which then slightly outperforms Zyzzyva. The comparison of rows $2$ and $3$ suggests that different protocols have different ``sweet spots'' depending on the request size and quorum size. When requests are small, optimistically waiting for $3f+1$ votes is reasonable, but when requests become larger, waiting for the slowest $f$ nodes to vote on a leader proposal takes a long time, especially when $f$ is large. In the latter scenario, protocols that only need $2f+1$ replicas to vote on the leader proposal perform better, even at the cost of an extra phase of exchanging small hashes and extra computation.
Note that due to the separation of transaction dissemination from consensus logic, only the leader proposals contain the actual requests.

Two of the protocols we studied require a $3f+1$ quorum in their fast paths: Zyzzyva and SBFT. Across rows 1 to 3 in Table~\ref{tbl:landscape}, Zyzzyva's performance leads or almost equals SBFT since it has fewer phases of communication. However, we found that with weaker clients, SBFT outperforms Zyzzyva in some cases. More specifically, we reran row $1$ on a different hardware setup, where the client machine (which hosts multiple client threads) is configured with fewer CPU cores and higher network latency: this is done by limiting the available CPU cores on a 10-core machine to 6 using \texttt{taskset} and injecting an extra $20$ms RTT. In this new setup, SBFT outperforms Zyzzyva by $8.5\%$. These two protocols demonstrate a design trade-off between the number of phases and the choice of commit collector. Since SBFT moves the collector role from clients to certain replicas, it is beneficial when clients have weak network connectivity and compute power. 

\noindent {\bf Row 4.}
Row 4 demonstrates the effect of non-responsive replicas, referred to as absentees in the table. When certain replicas are non-responsive, the performance of dual-path protocols is adversely impacted, since a slow path is activated only after a timer expiration when failing to gather $3f+1$ votes on the fast path. Conversely, single-path protocols are less impacted and even have better performance due to fewer message validations and less bookkeeping. In this setting, CheapBFT is the best-performing protocol in our evaluation, which outperforms the next best protocol HotStuff-2 by $13.1\%$ since it has fewer phases of communication.
Due to their slow paths, Zyzzyva and SBFT become the bottom-performing protocols.
It is worth mentioning that in our evaluation, HotStuff-2 is equipped with a reputation-based leader rotation mechanism, Carousel~\cite{cohen2022aware}, which tracks active replica participation via their signed votes during the committed chain prefix in order to select the next leaders. For a HotStuff-2 implementation without leader reputation mechanisms, CheapBFT will outperform it by an even larger margin, since it suffers from a non-responsive leader periodically.

\noindent {\bf Row 5-8.}
Row 5-8 evaluate different degrees of proposal slowness, representing a Byzantine world where malicious leaders might deliberately slow down the system. When the slowness is as low as 20ms and $f$ is large, HotStuff-2 outperforms all other protocols by up to $45.4\%$. This is due to HotStuff-2's \textit{routine} leader rotation, which is made possible with low cost by its linear responsive view-change and alleviates the impact of slow nodes elected as leaders. Although Prime also replaces slow leaders proactively, it has more phases and higher communication complexity than HotStuff-2 (i.e., $6$ phases and quadratic complexity compared to $2$ phases and linear complexity), resulting in worse performance. However, when the slowness further increases to $100$ms, or when the network size reduces from $13$ to $4$ ($f=4$ to $f=1$), the ranking flips: Prime outperforms HotStuff-2 by $16.9\%$-$71.5\%$. This happens since Prime replaces any deliberately slow leader with a \textit{stable} benign leader. In Prime, each node measures the actual turnaround time to the leader, which is independent of the system load, and compares it with the acceptable turnaround time, which is a function of the RTT between correct servers. In HotStuff-2, when the network is small or slowness is high, slow nodes are being rotated in as leaders routinely, causing considerable slowness. This outweighs the benefits of HotStuff-2's simpler and linear communication pattern, resulting in sub-optimal performance.

These experiments demonstrate a complex trade-off between design principles of different BFT protocols. Thus, when conditions change dynamically, no single protocol outperforms others in all scenarios.

\subsection{The Case for Reinforcement Learning}

Since no single BFT protocol is dominant in all scenarios, one could imagine building heuristics or supervised learning models that map conditions to the best-performing protocol, and switching protocols at run-time according to the current perceived conditions. However, such approaches suffer from several drawbacks.

{\noindent \textbf{Condition space size.} The condition space is too large to search. Table~\ref{tbl:landscape} only presents a sample of the space. The complete condition space we monitor consists of 6 dimensions (State 1 and 2 in Section~\ref{sec:state_action}), where each dimension is either a continuous or discrete variable. Further, each point in the condition space has multiple protocols to experiment with. Even with coarse-grained sampling and an automated toolkit, it took us more than a week to experimentally explore just a small subset of the condition space. Unfortunately, building good heuristics and supervised models requires \textit{complete} data, which are hard to obtain.}
    
{\noindent \textbf{Hardware and time dependence.} The mapping from conditions to the best-performing protocol depends on the underlying hardware and system configuration: when changing from \texttt{xl170} to \texttt{m510} instances on CloudLab, or changing from a $13$-node network to a $4$-node network, the mapping changes. Additionally, we observe that even with the same hardware instance type but across different launches on shared facilities like CloudLab, the sweet spots (in terms of request size) of Zyzzyva and CheapBFT change due to subtle differences in physical machines and network resource availability. Worse yet, network conditions can change over time, rendering any pre-computed mapping on a specific network less useful.}
    
{\noindent \textbf{Growing protocol space.} When new BFT protocols emerge (e.g., HotStuff-2) or changes are introduced to existing implementations (e.g., DiemBFT-v1 to v4~\cite{diembft-v4}), any precomputed mapping would need to be recomputed. In other words, one would need to re-collect data and re-craft the heuristics, or re-collect data and retrain the supervised model virtually every time a new BFT consensus protocol is proposed.}

Reinforcement learning addresses this daunting and complex task, and has shown superior performance in other learned systems~\cite{bao, decima, learned_gc, wu2022adachain}. Unlike supervised learning, which assumes training data is complete and requires a separate lengthy data collection process prior to deployment, one can simply plug and play an RL-based system --- it learns from its mistakes and optimizes long-term rewards through trials in an online fashion. With reinforcement learning, \sys can optimize itself to whatever client workloads, faults, hardware, system configurations and BFT protocols present, providing adaptivity and significant operational benefits. 

%% file: sec_overview.tex
\section{\sys Overview} \label{sec:overview}

We provide an overview of \sys, first outlining its system model followed by its overall reinforcement learning based design.

\subsection{System Model} \label{sec:trust}

In \sys, we assume a system consists of a fixed set of $n=3f+1$ \textit{nodes} and a finite population of \textit{clients}, where up to $f$ nodes and any number of clients might experience Byzantine faults. Each node serves two roles simultaneously, \textit{validator} and \textit{learning agent}. A validator is responsible for totally ordering the blocks, while a learning agent coordinates online data collection, trains machine learning models, and periodically instructs the companion validator at run-time to replace the current BFT protocol.

When a node is faulty, it can behave arbitrarily in any of its roles. The faulty validators may exhibit standard malicious behavior such as double voting and message suppression or equivocation. We assume a strong adversary that can coordinate the faulty validators to compromise the replicated service. However, we do assume the adversary cannot break the cryptographic techniques. Given our use of machine learning, faulty learning agents could exhibit additional malicious behaviors related to learning, such as being non-responsive (e.g., refuse to exchange its locally measured data with other learning agents), as well as presenting \textit{arbitrarily manipulated} local data points (i.e., local features and rewards) in order to disrupt the machine learning models on other agents.

Network communication is point-to-point and authenticated. \sys adopts the \textit{partial synchrony model}~\cite{dwork1988consensus}, where there is a known bound $\Delta$ and an unknown Global Stabilization Time (GST), such that after GST, all transmissions between two correct nodes arrive within time $\Delta$. For two different roles on the same node, we assume their communication in-between is always synchronous.

\subsection{Design Overview} \label{sec:design}

At a high level, \sys contains three key components: (1) a reinforcement learning algorithm (i.e., the core of the learning agent) that guides the choice of BFT protocols according to the perceived underlying dynamic environment, (2) a coordination protocol that collects data distributedly at run-time, and (3) a switching mechanism that allows \sys to seamlessly transition from one BFT protocol to another while ensuring safety and liveness.

\sys operates in \textit{epochs}, where each epoch is marked by the completion of $k$  blocks. Here, $k$ is a predefined constant hyper-parameter. Within one epoch, the protocol remains unchanged. When the learning agent finds a protocol candidate, it instructs the validator to use that protocol for the next epoch. We next introduce \sys's key components and workflow.

\sparagraph{Learning agent.}
\sys's learning agent models the problem of selecting a BFT protocol as a contextual multi-armed bandit (CMAB) problem~\cite{bandit_survey}: periodically, \sys examines the most recent state of the workload and faults in the system (\emph{context}), and then selects one of many BFT protocols (\emph{arms}) in our protocol pool. After making the selection, it observes the performance of the newly selected protocol (\emph{reward}). To be successful, \sys must balance the \emph{exploration} of new, untested protocols with \emph{exploiting} past experience to maximize performance. That is, without a careful balance of exploration and exploitation, \sys risks failing to discover an optimal protocol (too much exploitation), or performing no better than random (too much exploration). We select this CMAB formulation (as opposed to full reinforcement learning) because CMABs are exceptionally well-studied, enable faster convergence, and many asymptotically-optimal algorithms exist to solve them~\cite{thompson_intro, thompson_bound}. Details about the learning algorithms are provided in Section~\ref{sec:learningalgo}.

Since \sys operates in a Byzantine environment, a centralized learning agent cannot be trusted. In \sys, each validator process has a companion learning agent running on the same node, and accepts instructions only from its companion learning agent. The learning agents themselves also form a state machine replication. Specifically, they start with the same \textit{initial state}, i.e., the same random seed of machine learning models. For the same epoch $t$, as we will show later, different learning agents agree on the same \textit{sequence of operations}, i.e., training data points where each data point consists of context and reward. 
With deterministic training, benign learning agents host the same parameters for their machine learning models. As a result, if different learning agents perceive the same context for epoch $t+1$, they will render the same decision (i.e., choice of protocol) for epoch $t+1$.

\sparagraph{Distributed online data collection.}
In a Byzantine environment, no centralized entity could be trusted to collect training data. Therefore, the learning agents in \sys also participate in a protocol that coordinates distributed data collection in an online fashion. At a high level, for every epoch, each learning agent monitors its local context and reward at runtime, then exchanges them with other agents via a separate instance of BFT consensus independent of the consensus that validators are running. 
For each epoch, agents form agreement over an aggregation of contexts and rewards that include input from at least a quorum of two-thirds of agents. The consensus algorithm used for forming this agreement is left open for the system designer (note that it is invoked only once per epoch, hence does not need to have high throughput).
Once an agreed quorum of local contexts and rewards is obtained, each learning agent can apply the same robustness filter to the quorum in order to get a global context and reward, constituting a training data point. Details about this learning-coordination protocol are provided in Section~\ref{sec:collect}.

\sparagraph{Switching BFT protocols.}
After a BFT protocol is selected by the learning agent, the switching mechanism allows each validator to make use of this protocol for the next epoch. Our switching mechanism is an improvement over \abs~\cite{aublin2015next}, which aborts a BFT instance if a certain progress condition is not met. An epoch in \sys is equivalent to a \textit{Backup} instance in \abs. Due to space limits, details on how we switch BFT protocols are provided in Appendix~\ref{sec:appendix_switching}.

\begin{figure}[t]
\centering
\includegraphics[width=\linewidth]{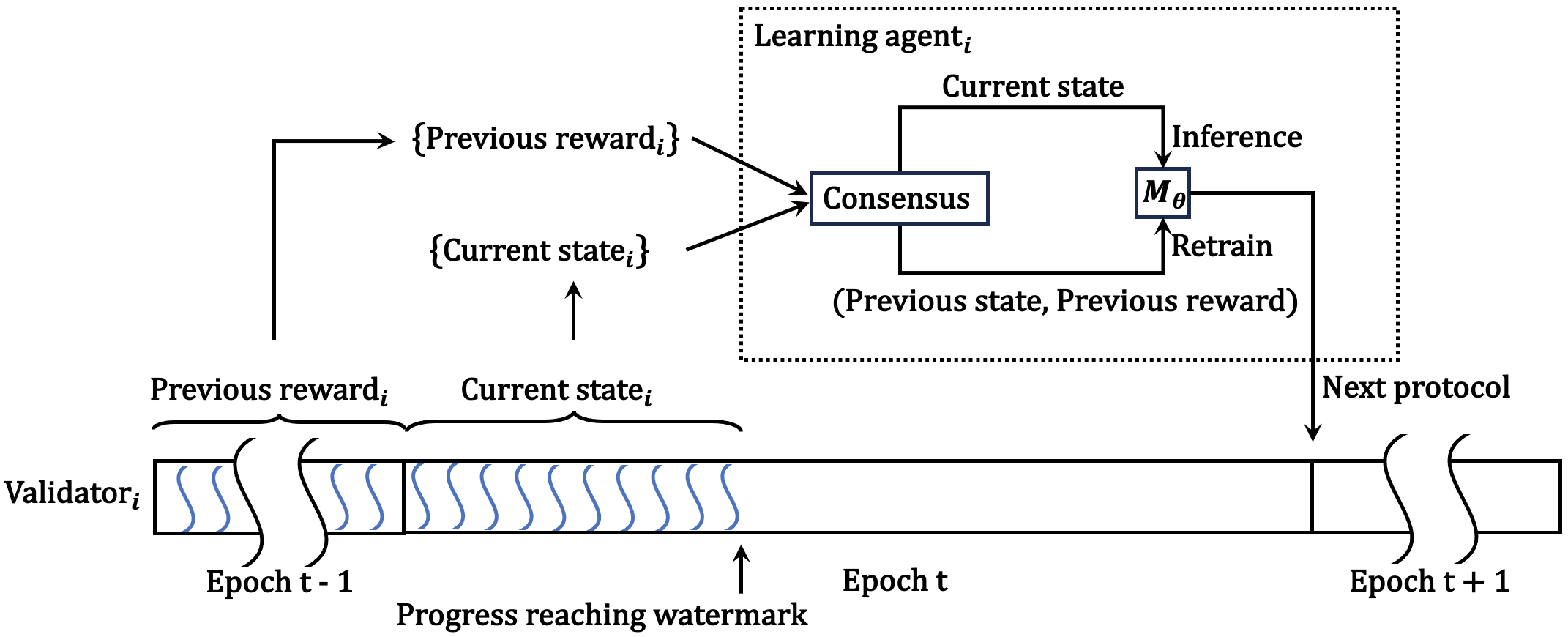}
\vspace{-1em}
\caption{Overview of \sys. For readability, we only present the internals of one node $i$.}
\vspace{1em}
\label{fig:overview}
\end{figure}

\sparagraph{\sys workflow overview.}
Figure~\ref{fig:overview} presents an overview of \sys, where each node in the system follows the same workflow as depicted. In the middle of epoch $t$, when the number of executed blocks reaches a certain watermark, the validator on node $i$ notifies its local learning agent. The learning agent featurizes its current local state (i.e., context) observed in epoch $t$, and uses it to approximate the next local state $state_i^{t+1}$ for epoch $t+1$. Each agent exchanges $state_i^{t+1}$ and its locally measured reward of previous epoch $reward_i^{t-1}$ with other agents via the learning-coordination protocol. Therefore, each agent obtains the same global state $state^{t+1}$ and reward $reward^{t-1}$. Subsequently, each agent adds the $(state^{t-1}, protocol^{t-1}, reward^{t-1})$ triplet to its experience buffer, and retrains its predictive model $M_{\theta}$ based on its experience buffer as well as the chosen algorithm to solve CMAB. Once retrained, the predictive model $M_\theta$ inferences the performance of each protocol candidate under $state^{t+1}$, and selects $protocol^{t+1}$ that is predicted to have the best performance. The learning agent then informs the validator to switch to $protocol^{t+1}$ for epoch $t+1$, the reward of which is then measured locally upon reaching the end of epoch $t+1$. The validator only starts epoch $t+1$ once it receives a decision for that epoch from the companion agent.

The learning agent is designed in a way such that the BFT system is not delayed due to learning. 
First, within one epoch, when the model undergoes retraining and inference, the parallel validator process still commits blocks simultaneously. Second, with a lightweight model design and limiting the size of the experience buffer, model training and inference can be viewed as a synchronous process. In other words, with a reasonable epoch length, the learning agent can complete protocol selection before the validator finishes its current epoch, without impeding the start of the next epoch.

%% file: sec_learning.tex
\section{Learning Algorithms} \label{sec:learningalgo}

This section delves into \sys's learning algorithms. We first formalize the learning problem and explain the use of Thompson sampling. The state and action space design is then outlined, followed by the predictive model description.

\subsection{Problem Formulation}

\sys formulates the learning problem as a contextual multi-armed bandit (CMAB) problem, where an agent periodically makes decisions in a sequence of epochs. In epoch $t$, the agent selects an action $a_t$ in its protocol pool based on a provided state $s_t$, and then receives a reward $r_t$. The agent's goal is to select actions in a way that minimizes \textit{regret}, i.e., the difference between the reward sum associated with an optimal selection strategy and the reward sum associated with the chosen actions. CMABs assume that epochs are independent from each other, and that the optimal action depends only on the state $s_t$. \sys fits this assumption, since the current choice of protocol does not affect the pattern of workloads and faults in future epochs. Using this CMAB formulation, \sys allows users to specify any performance metric (e.g., throughput or latency) as the reward function to optimize.

\sparagraph{Thompson sampling.} 
Amongst different CMAB algorithms, we select Thompson sampling~\cite{thompson_intro, thompson_bound} for its simplicity: at the start of each epoch, the learning agent trains a model based on current experience, and then selects the best action as predicted by the model. 
In Thompson sampling, instead of selecting the model parameters that are most likely given the training data as used by supervised learning, it \emph{samples model parameters proportionally to their likelihood given the training data}. More formally, we can define maximum likelihood estimation as finding the model parameters $\theta$ that maximize likelihood given experience $E$: $\argmax_\theta P(\theta \mid E)$ (assuming a uniform prior). Instead of maximizing likelihood, Thompson sampling simply samples from the distribution $P(\theta \mid E)$.  As a result, if we have a lot of data suggesting that our model weights should be in a certain part of the parameter space, our sampled parameters are likely to be in that part of the space. Conversely, if we have only a small amount of data suggesting that our model weights should be in a certain part of the parameter space, we may or may not sample parameters in that part of the space during any given epoch. 

\subsection{State and Action Space} \label{sec:state_action}
We next list factors that affect the performance of BFT protocols, broadly grouped into workloads, faults, and hardware/system configurations categories, jointly constituting the state space. Within each epoch, each learning agent leverages a window of the last $w$ executed requests to featurize such factors, where $w$ is a constant hyper-parameter.



\noindent \textbf{State 1: Workloads (W).} 
The first category consists of factors that are influenced by application and client dynamics.

\noindent \underline{\textit{W1: Request size.}} The request size is dependent on the application workload, where some requests contain little data while others are more involved and require updating files with large chunks of data. Although all our candidate protocols separate request dissemination from sequencing 
(i.e., only the leader proposals contain the actual requests while the remaining messages contain the hash of requests), as described in Section~\ref{sec:landscape}, request size is still an important factor impacting the performance of different protocols in different ways. We use the average request size to represent this feature.

\noindent \underline{\textit{W2: Reply size.}} Depending on the application, request and reply size can be asymmetric. Reply size also impacts different protocols in different ways, but with a distinct boundary from that of request size. For instance, in our experiments, when all other factors remain the same, CheapBFT performs best for either 0KB/4KB (request size/reply size) or 40KB/0KB, while Zyzzyva is best with 4KB/0KB. Furthermore, different protocols have different sensitivity to an increase in reply size. For instance, SBFT reduces the per-client linear reply cost to just one message by adding an execution aggregation phase. We use the average reply size to represent this feature.

\noindent \underline{\textit{W3: Load on system.}} The load on the BFT system is dependent on the number of clients and the rate at which clients send new requests. Specifically, each honest client allows a \textit{quota} of outstanding unacknowledged requests before issuing new ones, controlling the rate at which requests are generated relative to the system's capacity to process them. 

In our experiments, we observed that lowering the load affects different protocols in the same direction, but to different extents: for example, there is a sharper drop in the throughput of Zyzzyva and SBFT compared with other protocols. The reason is that lower load increases batching delay, which has a larger impact in terms of both latency and throughput on protocols with fewer phases (i.e., Zyzzyva and SBFT in our protocol pool). 
\sys derives the per-client sending rate according to the request timestamps
, and uses the aggregated client sending rate to represent this feature.

\noindent \textit{\underline{W4: Execution overhead.}} Execution overhead captures the computational cost of request execution, which impacts the system in two ways. First, it directly affects the execution latency in state machine replication. Second, it indirectly affects other components of the BFT protocol that are also compute-intensive. For instance, requests with high execution overhead compete for CPU resources that are otherwise used to sign and verify messages, especially when machines have limited compute capacity or a small number of cores. Higher compute load results in excessive context switching, and potentially pushes the system towards being compute-bound instead of network-bound. We use the CPU cycles consumed by the executor thread to represent this feature.

\noindent \textbf{State 2: Faults (F).}  
The next category of factors is tied to faulty behaviors. BFT protocols make different assumptions about ``steady state'' and ``common faults'', and hence, each protocol is often optimized for specific fault scenarios. The features below enable \sys to tell what type of fault scenarios the system is experiencing and choose the most promising protocol accordingly. 

Note that \sys does not aim to defend against \textit{transient} or \textit{broad-spectrum} faults. The reason is, transient faults are already handled timely at the protocol level and require no protocol switching.
An example would be a crashed leader or a malicious leader which equivocates, such that no progress is made and a view-change will be triggered to replace the leader. For broad-spectrum faults, effective and orthogonal solutions already exist. Examples of broad-spectrum faults include network flooding which can be resolved via resource isolation~\cite{aublin2015next}, and malformed client requests which can be handled by enforcing client signatures (instead of MACs)~\cite{clement2009making}.

\noindent \textit{\underline{F1: Absence from participation.}} 
In BFT consensus, validators can be absent from participation for various reasons: a (benignly) crashed validator is absent from all protocol phases after crashing, while an alive malicious validator could be absent from any arbitrary phases. 

Measuring absence is tricky in a Byzantine environment, especially considering that a collusion of $f$ malicious participants could taint validator participation simply by \textit{excluding} some alive benign validators (up to $f$) and progressing without them. For example, a malicious leader could deliberately avoid sending leader proposals to them, while the $f$ malicious validators work together with the remaining $f+1$ benign ones to make sure all requests are committed successfully on these $2f+1$ validators. We refer to such excluded, non-faulty validators as being placed \textit{in-dark}. In-dark validators could further be excluded in other protocol phases, in addition to leader proposals, by $f$ malicious validators. Since no state transition is ever triggered on in-dark validators, they remain in the initial state and are thus absent from participation. Although they will timeout and complain, since fewer than $f+1$ validators complain, view-change is not triggered to replace the malicious leader and they are in-dark continuously.

All protocols tolerate the absence of up to $f$ validators by design, but the performance of different protocols is impacted differently. As illustrated in Section~\ref{sec:landscape}, dual-path protocols (i.e., Zyzzyva, SBFT) are adversely impacted since the more expensive slow paths are initiated, while single-path protocols (i.e., PBFT, CheapBFT, Prime, HotStuff-2) could be positively impacted due to less resource consumption. 


The learning agents in \sys featurize absence by utilizing information that is already collected locally during protocol execution. First, \textit{fast path ratio} captures the percentage of slots committed in the fast path over the total number of committed slots. For single-path protocols, all slots are committed in the slow path. Second, for each slot, the agent sums the number of (valid) distinct messages 
from each sender, deriving the \textit{number of received messages per slot}.
Note that this feature does not incur more messages to be sent or received; it simply counts messages as they arrive and pass preliminary processing (de-serializing and sender verification) before they can be excluded from protocol steps like voting. 

\noindent \textit{\underline{F2: Slowness of proposal.}} 
In leader-based BFT protocols, every slot is initiated by a leader proposal, which significantly affects the system's end-to-end performance. In the case of a faulty leader, validators use a timer to trigger view-change, which will replace the leader, hence guaranteeing liveness. However, a malicious leader can deliberately slow down its proposals without being replaced, resulting in poor latency and throughput. 
Slowness may not necessarily be a deliberate act by a malicious leader, it could result simply from a weak or overloaded leader, albeit to a lighter extent. 

Section~\ref{sec:landscape} illustrates the effect of this factor. Protocols with routine or proactive leader replacement (i.e., HotStuff-2, Prime) maintain good performance under such slowness, but perform sub-optimally in other normal cases. To featurize this factor, each node in \sys timestamps every leader proposal received, and measures the average time interval between receiving two consecutive proposals.


\noindent \textbf {State 3: Hardware and system configurations.} 
The last category comprises hardware and system configurations. Hardware-level factors include standard data-center infrastructure network settings that affect network latency and bandwidth, and also machine-level configurations such as CPU frequency and the number of cores. System-level configurations include the number of nodes in the consensus system and the geo-distribution of the nodes. Compared to State 1 and 2, both hardware and system configurations are fairly static and do not change rapidly at the timescale of a consensus deployment. Thus, \sys does not explicitly featurize these factors because (1) the impact of these factors is implicitly encoded in the predictive model trained online, and (2) CMABs will converge even without any explicit features~\cite{nonstationary_thompson}, the purpose of which is to accelerate convergence so that the optimal action is reached before the world changes.

\noindent \textbf{Actions.} 
\sys's action space consists of the same set of leader-based protocols that are studied in Section~\ref{sec:landscape}, namely, PBFT, Zyzzyva, CheapBFT, Prime, SBFT, and HotStuff-2. As we will show  later, since \sys builds a separate model for each candidate protocol, features 
of the protocol design are encoded in the model itself. Thus, \sys does not require protocol-specific feature engineering for encoding the action space, making it easy to incorporate new protocols into the \sys framework. Although \sys focuses on choosing the best-performing protocol, configuring protocol internal parameters (e.g., values of different timers and the interval for rotating leaders) intelligently at run-time can be an interesting future extension. 

\subsection{Predictive Model}

Each learning agent hosts some predictive models, which follow the \textit{value based} RL approach: given the featurized current state, predict the performance (i.e., reward) of each action (i.e.,  protocol). The simplest implementation would be to build a single predictive model for all protocols, but this has a major drawback. While features \textit{W1-W4} in the workloads category are completely independent from the previous action, the featurized absence-from-participation $F1$ and slowness-of-proposal $F2$ have a ``one-step dependency'': the current observed $state_{F1, F2}^{t+1}$ is dependent on the previous $protocol^t$.  When the workload and fault scenario shift, such one-step dependency might prevent convergence to the new optimal protocol. For instance, if \sys has converged to protocols whose leader has lower parallelism (e.g., Prime due to message aggregation in global ordering), the measured slowness of proposals will be higher than other protocols, regardless of whether a fault is actually happening or not. In other words, the interpretation of the slowness-of-proposal feature changes based on the previous action. If the model learns that a high slowness-of-proposal is bad for Zyzzyva, then once Prime is chosen, slowness-of-proposal will always seem high, and Zyzzyva may not ever be selected again.

Luckily, unlike in the general case of Markov decision processes, the dependency observed here is limited to a single time step. That is, the prediction of the next best action is independent \emph{given the immediately prior action}. To solve this issue rooted in one-step dependency, the learning agent trains a separate model for each possible (previous protocol, protocol) pair, and divides the experience buffer into several smaller \textit{buckets} according to the (previous protocol, protocol) pair as well. In terms of bandit theory, assuming there are $K$ protocols in our action space, \sys's approach is equivalent to playing $K$ bandit games where each game has $K$ arms. In each game, the current observed $state_{F1, F2}^{t+1}$ is independent from the previous action  $protocol^t$.

It is worth mentioning that this transformation does not completely remove the one-step dependency. The action made at epoch $t$ will determine which of the $K$ bandit games is being played at epoch $t+1$. A multi-armed bandit algorithm will not be able to take advantage of the fact that some of the $K$ bandits may have significantly better reward distributions than others. Thus, the convergence bound on regret of bandit algorithms will not apply to our scenario. However, since each of the $K$ bandits will be played an unbounded number of times eventually (assuming the probability of any action never fully reaches 0), regret is still bounded in the limit. 


Specifically, for each possible $(protocol^t, protocol^{t+1})$ pair, \sys uses a lightweight random forest~\cite{rf} as the predictive model, which is trained exclusively on the corresponding experience bucket. The model takes the featurized state as input, and outputs the predicted performance for the corresponding candidate action $protocol^{t+1}$. Thus, at inference time, given a known previous protocol and the current state, the learning agent enumerates $K$ models to get the predicted performance for each candidate protocol, and then chooses the candidate with the best predicted performance to be carried out. Once there is a tie on the best predicted performance, we break the tie randomly to avoid local maxima. When an experience bucket is empty, \sys prioritizes exploring this bucket by choosing the corresponding candidate protocol to be carried out.

\sparagraph{Integration with Thompson sampling.}
Integrating a predictive model with Thompson sampling requires the ability to sample model parameters from $P(\theta\mid E)$ --- the distribution of model parameters given the current experience. The simplest technique (which has been shown to work well in practice~\cite{thompson_bootstrap}) is to train the model as usual, but only on a bootstrap~\cite{bootstrapping} of the training data. In other words, the predictive model is trained using $|E|$ random samples drawn with replacement from experience $E$, inducing the desired sampling properties. \sys uses this bootstrapping technique on each experience bucket and predictive model for its simplicity.

\sparagraph{Overhead of learning.}
First, \sys's training overhead is not larger than the strawman of building a single predictive model, since in every epoch, only one model which corresponds to the updated bucket needs to be retrained. For such a bucket, the time complexity for training a single random forest is $O(n\log n)$, where $n$ is the number of data points. Thus, given the same total population of data, it even incurs less training overhead than the strawman solution, since the bucket contains fewer data points than the single unified experience buffer. Second, \sys's inference overhead is $O(K)$, where $K$ is the number of candidate protocols. Lastly, \sys has the same memory overhead as the strawman for storing training data. However, it incurs $O(K^2)$ memory overhead for storing the models. Since random forest is a very lightweight model as compared to deep neural networks, such model storage overhead is negligible.

%% file: sec_collect.tex
\section{Learning Coordination} \label{sec:collect}
The goal of the \textit{learning-coordination} mechanism is to form an agreement at each epoch on a \textit{report quorum} that includes local metrics collected from $2f+1$ nodes.

Specifically, learning coordination is performed in every epoch $t$. After executing $w$ requests (a hyper-parameter) in epoch $t$, each node $i$ gathers local performance indicators $p_i^{t-1}$ measured during epoch $t-1$, featurizes the next state $f_i^{t+1}$, and broadcasts both metrics inside a \rpt message. 
To ensure that at least $f+1$ metrics in the report quorum are honest measurements, it is important that the metrics reported by honest nodes are measured by \textbf{themselves}. That is, if a node $j$ has been placed in-dark (defined in Section~\ref{sec:state_action}) or temporarily slows down during epoch $t$, it may not have executed $w$ requests by itself. Rather, node $j$ will have recovered the consensus state through a \textit{state-transfer} from other nodes. In this case, $j$ should avoid reporting the state features it has copied from others, and likewise, avoid reporting performance indicators collected from partial or no execution. Therefore, node $j$ will not report any metrics for epoch $t$. Note that in addition to the $f$ benign nodes being placed in-dark, in the meantime, the $f$ Byzantine nodes that contributed to committing requests can refuse to report their metrics. Hence, there may not be enough $2f+1$ nodes reporting for the epoch.



In order for nodes to agree on a quorum of (valid) reports to be used as input for the learning engine, any ``blackbox'' validated Byzantine consensus primitive (VBC) seeded with a leader collecting reports from $2f+1$ nodes can be utilized. Specifically, for each epoch $t$, the leader of VBC initiates {\small \textsf{VBC-PROPOSE}}(($t, reportQC^t$), $P$) once it receives valid \rpt messages ($p_i^{t-1}$,  $f_i^{t+1}$) where both fields are non-null from $2f+1$ nodes, or when a timer expires. Here, $P$ is an external validity predicate that checks if $reportQC^t$ includes at least 
$f+1$ distinct reports. 

Each node participating in VBC gates voting for a leader proposal it receives by applying the validity predicate $P$ to it. Once a quorum of reports $reportQC^t$ is decided by VBC, if it includes sufficient $2f+1$ reports, each node takes the median value of each field in order to obtain a robust global performance measurement $p^{t-1}$ and state feature $f^{t+1}$, thereby triggering the retraining and inference process. Taking the median value from an aggregated set of metrics guarantees that despite $f$ arbitrarily manipulated values from Byzantine nodes, the global value taken is between two honest measurement values. Otherwise, if $reportQC^t$ does not include sufficient reports, each node retains the decision from the previous epoch instead of deriving any new learning decision, and complains about the leader in VBC as well as the leader in the current protocol used by the node for committing client requests. Note that since VBC is a separate consensus instance, the leader of VBC can be different from the leader of the current protocol in \sys. Either of them acting maliciously can result in insufficient reports being collected.


For completeness, Appendix~\ref{sec:appendix_protocol} presents in Algorithm~\ref{alg:collect} a detailed learning coordination protocol where VBC is implemented using PBFT. Due to space limitations, the safety, liveness, and robustness guarantees of \sys's learning agents are discussed and proved in Appendix~\ref{sec:appendix_proof}.

%% file: sec_impl.tex
\section{Implementation} \label{sec:impl}

We implemented \sys in Java using Bedrock \cite{amiri2024bedrock}, a unified platform for BFT protocol implementation and experimentation. The Bedrock platform consists of four main components: the core unit, the state machine manager, the plugin manager, and the coordination unit. The core unit defines entities (e.g., clients and validators), maintains the application logic, enables users to specify different workloads and benchmarks,
and track the execution of requests via sequence numbers and views. The state machine manager parses Bedrock's domain-specific language (DSL) for rapidly prototyping BFT protocols, and defines the states and transitions of the specific BFT protocol for each entity. The plugin manager enables implementing protocol-specific behaviors that cannot be captured by Bedrock's DSL, while the coordination unit manages the run-time execution of Bedrock.

All protocol candidates in \sys's action space utilize Bedrock. On top of it, we implemented a workload and fault generator, which allows users to specify time-varying dynamics inside a {\tt YAML} configuration file, enabling parameterized random sampling and scheduling of predefined sequences of events. To enable switching among BFT protocols, we implemented a new state machine manager which loads all plugins required by \sys when the system boots, and uniquely tags each protocol state and transition such that different protocols do not interfere. This approach has negligible overhead since such loading and tagging only happens when the system boots, and lookup for states and transitions at run-time takes a (small) constant time. We implemented the learning agent separately in Python using {\tt scikit-learn}, which communicates with the companion validator process in Bedrock via gRPC. Our implementation is publicly available\footnote{\url{https://github.com/JeffersonQin/BFTBrain}}.

%% file: sec_eval.tex
\section{Evaluation}\label{sec:eval}

Our evaluation aims to answer the following questions: 
\begin{enumerate}[parsep=0.5mm, leftmargin=0em,labelwidth=*,align=left]
\item How fast can \sys converge to the best-performing protocol 
under static conditions without pre-training? 
\item How does \sys compare to fixed protocols and existing learning-based approaches in dynamic environments? 
\item How does the hardware setup affect the performance of fixed protocols? How does \sys compare to existing learning approaches under different hardware setups? 
\item How robust is \sys and how does it compare to existing learning approaches under adversarial data pollution? 
\item What learning overhead does \sys introduce? 
\end{enumerate}

In the rest of this section, we present the experimental setup and then answer each of the above questions.

\subsection{Experimental Setup}\label{sec:exp_setup}

\noindent \textbf{Testbed.} Unless otherwise specified, our testbed consists of several bare-metal xl170 machines on CloudLab, each with a 10-core Intel E5-2640v4 processor at 2.4 GHz, 64GB ECC Memory, and two dual-port Mellanox ConnectX-4 25 GB NIC. Each server is connected via a 10Gbps control link and a 25Gbps experimental link to Mellanox 2410 switches in groups of 40 servers. Our experiments are conducted using the experimental link.

\noindent \textbf{System configuration.}
Our experiments are carried over networks of two sizes: $n=4$ and $n=13$. 
In all experiments, we run multiple client threads on a separate xl170 server, where each client follows the standard closed-loop buffer design, i.e., allows at most 100 outstanding unacknowledged requests before issuing new ones. The number of clients is a parameter in our workload space. We use batch sizes of 10 requests throughout the experiments. Every experiment of \sys starts with PBFT as its initial protocol. We use throughput as the reward function for \sys to optimize.

\subsection{Convergence under Static Conditions}\label{sec:exp_static}

\begin{table*}[t]
\caption{Throughput of protocols and the convergence time of \sys under various static conditions. The highest throughput in each row is highlighted in {\color{blue}\textbf{blue}}.}
\vspace{0.5em}
\centering
\scriptsize
\begin{tabular}{l|lllllll|l}
\toprule
\multirow{2}{*}{Condition} & \multicolumn{7}{c|}{Average Throughput (tps)} & \multirow{2}{*}{Conv. Time (minutes)} \\ 
& PBFT & Zyzzyva & CheapBFT & Prime & SBFT & HotStuff-2 & \sys \\
\midrule
Row 1 (LAN) & 9133 & {\color{blue}\textbf{13664}} & 11822 & 4601 & 11067 & 6882 & 13100 & 0.81 \\
Row 4 (LAN)*& 10303 & 1025 & {\color{blue}\textbf{12297}} & 3749 & 2920 & 5156 & 11803 & 2.08 \\
Row 8 (LAN) & 989 & 988 & 989 & {\color{blue}\textbf{4527}} & 989 & 2640 & 4329 & 5.39 \\
Row 1 (WAN) & 5325 & 9503 & {\color{blue}\textbf{12201}} & 1639 & 8261 & 2882 & 11101 & 1.58 \\
\midrule
Average & 6438 & 6295 & 9327 & 3629 & 5809 & 4390 & {\color{blue}\textbf{10083}} & 2.47 \\
Worst   & 989 & 988 & 989 & 1639 & 989 & 2640 & {\color{blue}\textbf{4329}} & 5.39 \\
\bottomrule
\end{tabular}

\label{tbl:static}
\end{table*}

Our first set of experiments evaluates how quickly can \sys converge to the best-performing protocol under static conditions.
We picked three representative settings from Table~\ref{tbl:landscape} where the size of the network is small: row 1, a variant of row 4 where $f=1$, and row 8. Under each setting, we ran all six fixed protocols and \sys for $10$ minutes on a LAN. We note that the other rows in Table~\ref{tbl:landscape} yielded similar conclusions and are omitted for brevity.

Table~\ref{tbl:static} summarizes our results, listing for each static condition the average throughput (tps) of each protocol in the last $20$ epochs. The last column reports the convergence time of \sys, defined as the time \sys spent to reach the stable peak throughput. Although each setting has a different best-performing protocol, \sys always learns to select it within $0.81$-$5.39$ minutes\footnote{Most RL systems quantify convergence time in minutes.}.
%
As we will show soon, the convergence is significantly accelerated the second time \sys encounters the same conditions. 

Obviously, in this experiment, \sys is not able to surpass the best-performing protocol in any given setting. Additionally, as shown in the table, it experiences a marginal increase in switching overhead when compared to the leading protocol in each specific setting. Nevertheless, owing to its adaptability, \sys does deliver the best average and best worst-case performance across all settings in the table. In contrast, fixed protocols exhibit subpar average and worst-case throughput, underscoring our initial premise that no single fixed protocol excels under all circumstances.


\subsection{Adaptivity under Changing Conditions}\label{sec:exp_changing}
\begin{figure}[t]
\centering
\includegraphics[width=0.85\linewidth]{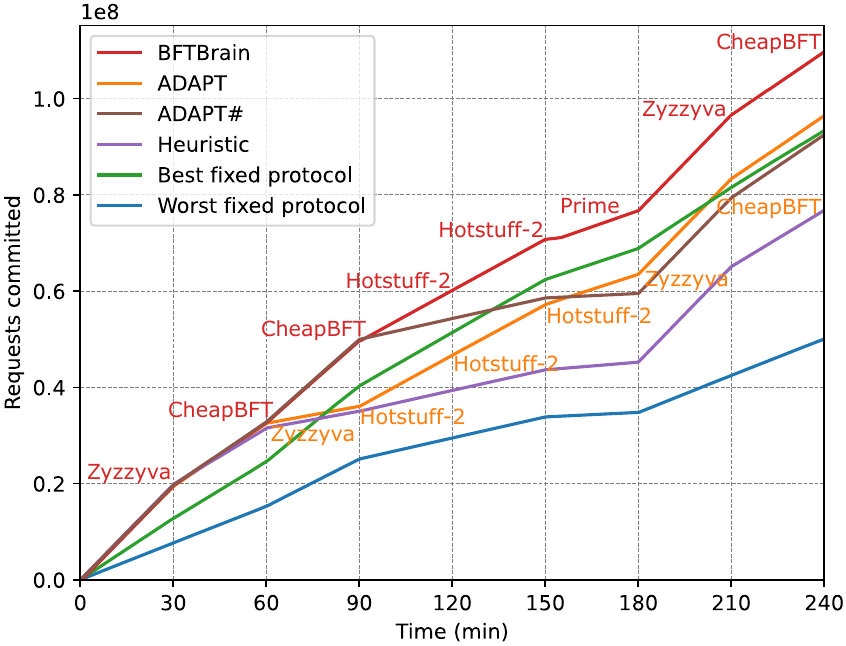}
\vspace{-1em}
\caption{Adaptivity of \sys under changing conditions. The vertical dashed lines indicate when conditions change. Labels indicate the dominant protocol that \sys and \adapt choose under each condition.}
\vspace{1em}
\label{fig:cycleback_cdf}
\end{figure}

\begin{figure}[t]
\centering
\includegraphics[width=\linewidth]{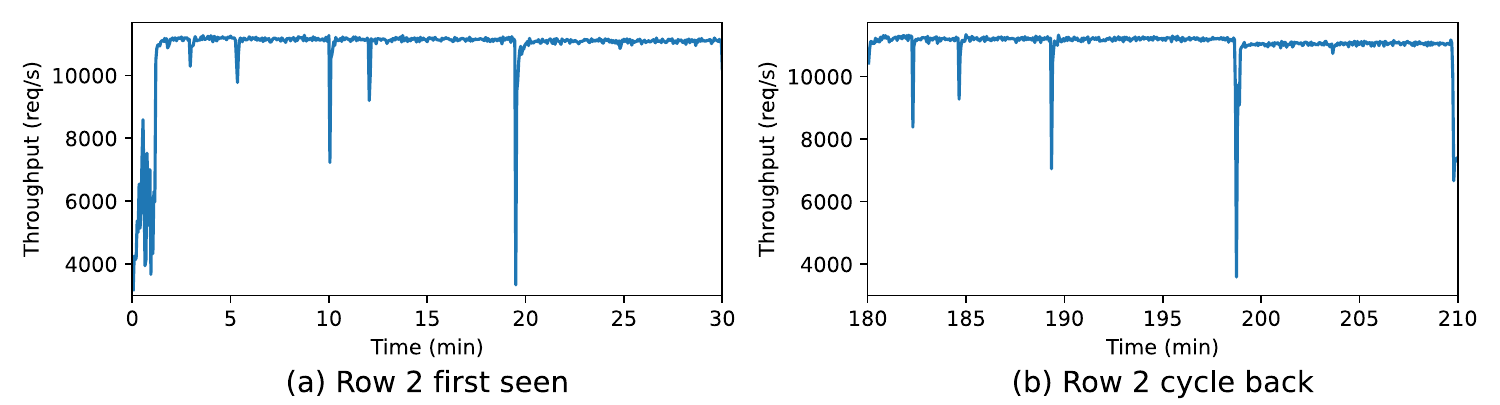}
\vspace{-2em}
\caption{\sys's throughput (a) during minutes $0$-$30$, and (b) during minutes $180$-$210$. In both periods, it encounters the system conditions captured in row $2$ of Table \ref{tbl:landscape}.}
\vspace{1em}
\label{fig:cycleback_pdf}
\end{figure}


Our next set of experiments evaluate \sys under dynamic conditions, demonstrating clear performance benefits. 


\sparagraph{Cycle back conditions.} In the initial series of experiments, we selected rows $2$-$7$ from Table~\ref{tbl:landscape}, each characterized by an identical network size ($f=4$). We ran the settings of each row for $30$ minutes each, employing a round-robin approach to switch to the next row and repeating the cycle from the beginning after reaching row $7$. We compared \sys against five baselines: the fixed protocols with the best and worst performance (based on the number of committed requests throughout the entire experiment), \adapt, \adaptsharp, and an expert heuristic. Details on the last three are explained below.

To be faithful to \adapt's design~\cite{bahsoun2015making}, we excluded all features capturing faults (i.e., State 2 in Section~\ref{sec:state_action}), and pre-trained \adapt with complete data that we collected in these changing conditions when running \sys for hours spanning multiple protocols. To study the effect of \textit{unseen conditions} in supervised learning, we implemented another baseline named \adaptsharp. In \adaptsharp, we used the same set of complete features as \sys, but pre-trained it on partial data that we collected in these changing conditions, i.e., excluding data corresponding to the settings of rows 5-7. In both \adapt and \adaptsharp, for fair comparison, we used the same set of BFT protocols as \sys in their action space. The expert heuristic we used is designed based on insights gleaned from Table~\ref{tbl:landscape}, namely: if proposal slowness is greater than 20ms, use Prime; otherwise, use Zyzzyva.

Figure~\ref{fig:cycleback_cdf} shows the cumulative number of committed requests over time, where the slope of each line indicates its current throughput. Throughout the $4$-hour experiment, \sys successfully converges to the best-performing protocols whenever the conditions change. In terms of the number of requests committed, \sys demonstrates $18\%$ improvement over the best fixed protocol (i.e., HotStuff-2), $119\%$ over the worst fixed protocol (i.e., PBFT), $14\%$ over \adapt, $19\%$ over \adaptsharp, and $43\%$ over heuristic.

As expected, both \adapt and \adaptsharp exhibit similar performance to \sys for the first $60$ minutes, as the system is operating without faults and their features and training data encompass these conditions. However, both methods encounter challenges when the conditions shift to rows $4$-$7$. \adapt struggles since it fails to recognize the changes in conditions, as faults are not captured in its feature space. While \adaptsharp does detect the changes due to its expanded feature set, these conditions were absent during its pre-training phase.  

Two things are worthy of noting. First, \sys not only outperforms both supervised learning based approaches, but also completely removes the cumbersome data collection and pre-training process prior to deployment. Second, in common stable conditions corresponding to row 2, \sys outperforms HotStuff-2 by $57\%$ while being able to switch to HotStuff-2 when conditions become advantageous for it.

Figure~\ref{fig:cycleback_pdf} further demonstrates how \sys's throughput changes over time: (a) during minutes $0$-$30$, and (b) during minutes $180$-$210$. In both periods, it encounters the conditions captured in row $2$ of Table \ref{tbl:landscape}.
Obviously, when the system condition cycles back to what it has seen before, \sys converges to the best-performing protocol much faster than the first time ($2$s vs. $70$s). 
The blips in both throughput plots indicate that certain sub-optimal protocols have been chosen for a few epochs. These are the explorations made by the Thompson sampling algorithm when it samples the less likely model parameters. When conditions are dynamic, such sampling is crucial to avoid being stuck at a local sub-optimal decision, since the predictive model can gain insights on the performance of unexplored protocols in the current condition.

\sparagraph{Randomized sampling.} 
Our next experiment evaluates \sys's adaptivity when there are \textit{more variations} in the state space and when the variations are \textit{more frequent}. Specifically, each dimension in the state space follows a certain normal distribution. We varied each dimension every $1$s by randomly sampling from its distribution, and shifted the mean and variance of such distribution every $20$ minutes. Appendix~\ref{sec:appendix_randomized} presents the detailed experiment setup and results. In a nutshell, during the entire $2$-hour deployment, \sys commits $44\%$ more requests than \adapt. This is a much larger improvement than the $14\%$ improvement demonstrated in the cycle back experiment and can be explained as follows. Certain input factors in the cycle back experiment are correlated, e.g., a request size near zero is correlated with high proposal slowness. Thus, although \adapt suffers from incomplete features, it indirectly learns the best-performing protocol under high proposal slowness using other features, as shown in Figure~\ref{fig:cycleback_cdf} between $90$-$150$ minutes. However, randomized sampling breaks such correlations, and as a result, \adapt does not have enough information to know how the surrounding environment changes which leads to poor performance.

\subsection{Adaptivity to Changing Hardware}\label{sec:exp_hardware}

The next set of experiments evaluate \sys under different hardware setups from our initial protocol evaluation, and compare with the adaptivity of \adapt to these new setups.
%
%
Specifically, we reran the settings of row $1$ from Table \ref{tbl:landscape} on a WAN (RTT=$38.7$ms, bandwidth=$559$Mbps) instead of on a LAN, where we placed two nodes in Cloudlab's Utah data center on xl170 machines and the other two nodes in the Wisconsin data center on c220g5 machines. We compared \sys with fixed protocols and with \adapt in this setup.

The results are summarized under ``Row 1 (WAN)'' in Table~\ref{tbl:static}; the detailed performance plot of \sys and \adapt is provided in Appendix~\ref{sec:appendix_wan}. Interestingly, compared to ``Row 1 (LAN)'' where Zyzzyva is the best protocol and leads by $15.6\%$, running the same workload and fault conditions on WAN with different instance types renders CheapBFT the best protocol, which outperforms Zyzzyva by $28.4\%$. The reason is that Zyzzyva commits via the fast path and requires votes from all nodes, resulting in high end-to-end latency, given the high RTT between some nodes. CheapBFT, on the other hand, requires a smaller quorum of voters, which could be co-located inside a single data center, thus achieving lower latency. With a moderate number of clients (i.e., $50$ in our workload), this results in CheapBFT also achieving higher throughput than Zyzzyva due to finite parallelism. Under this unseen hardware setup, \sys converges to the best-performing protocol within $1.58$ minutes without requiring data recollection prior to deployment. On the contrary, \adapt cannot transfer the knowledge it learned on different hardware setups. As a result, when \adapt is pre-trained with complete data we collected on ``Row 1 (LAN)'', it remains stuck at the sub-optimal decision Zyzzyva. 


This experiment demonstrates the operational benefits of \sys when hardware configurations may differ across deployments. \sys can rapidly converge to the best-performing protocol for new hardware settings without any human intervention. As discussed before, it may not be practical to recollect data and pre-train a supervised approach such as \adapt on every new hardware setup.

\subsection{Robustness of \sys}\label{sec:exp_robust}
\begin{figure}[t]
\centering
\includegraphics[width=0.85\linewidth]{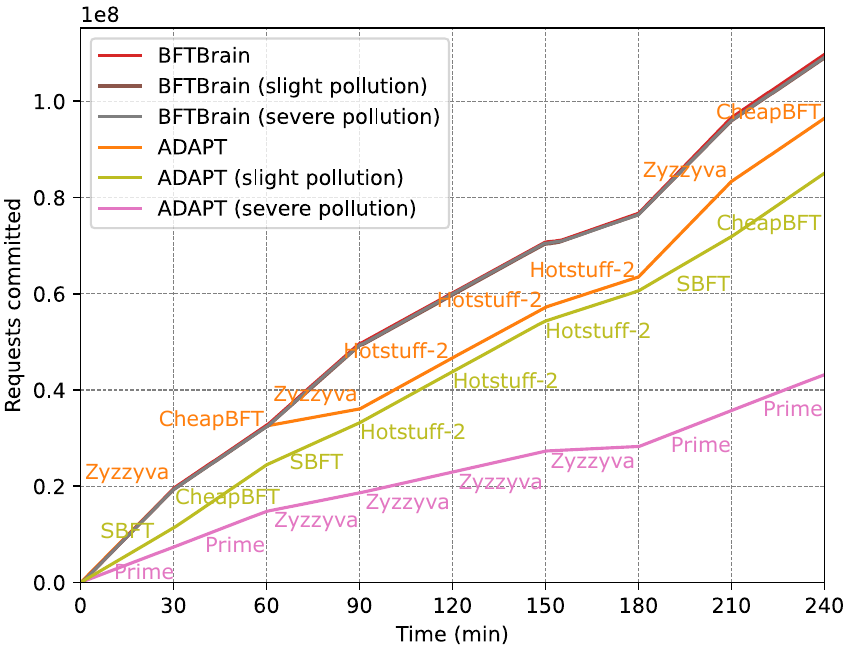}
\vspace{-1em}
\caption{Robustness of \sys against data pollution. The vertical dashed lines indicate when conditions change. Labels indicate the dominant protocol that \adapt (before and after pollution) chooses under each setting.}
\vspace{1em}
\label{fig:pollution}
\end{figure}

In a Byzantine environment, malicious nodes might arbitrarily manipulate (i.e., pollute) their collected metrics in order to misguide the machine learning models. Our next experiment compares the resilience of \sys and \adapt under data pollution.
We ran the same dynamic benchmark as the ``cycle back conditions'' experiment in Section~\ref{sec:exp_changing}, and compared \sys with \adapt's performance under two types of data pollution: slight and severe.  In the slight pollution scenario, only the reward (i.e., throughput) of SBFT was increased to 2.5x of its true value. In the severe pollution, regardless of the protocol, for every single data point, both its state and reward were polluted by replacing the true values with random values. The random value was chosen uniformly between 0 and 5x the maximal true value for the polluted dimension. Note that when polluting \sys, $f$ learning agents were malicious and hence reported polluted values as their local measurements; when polluting \adapt, the malicious centralized entity polluted training data and distributed the same data to all nodes in the system. 

Figure~\ref{fig:pollution} shows the number of committed requests with respect to time, under different types of pollution. Compared to non-polluted scenarios, under slight pollution, \sys incurs only a 0.7\% performance drop, while \adapt suffers from a 12\% drop. 
Thus, after such slight pollution where only less than 0.01\% of the data population is polluted, \sys outperforms \adapt by 28\%. Under severe pollution, although the $f$ malicious learning agents cause a distribution shift in \sys's global state and reward, \sys only incurs a 0.5\% performance drop. The reason is, as proved in Section~\ref{sec:appendix_proof}, the learning-coordination protocol guarantees that the global feature and reward always fall into the range between two honest local observations. On the contrary, when all features and rewards can be arbitrarily manipulated by the centralized entity in \adapt, it performs no better than randomly choosing protocols. In the worst case, a smart pollution strategy misguides \adapt to pick the worst protocol for each condition (as shown by the \adapt (severe pollution) line in Figure~\ref{fig:pollution}), resulting in a 55\% performance drop. In such a scenario, \sys outperforms \adapt by 154\%. 




\subsection{Overhead of \sys}\label{sec:exp_overhead}

Our last experiment evaluates the overhead incurred by \sys's learning framework. We repeated the ``cycle back conditions'' experiment in Section~\ref{sec:exp_changing}, and plot \sys's training and inference overhead in each epoch in Figure~\ref{fig:overhead} (provided in Appendix~\ref{sec:appendix_overhead}). Labels indicate the dominant protocol that \sys chooses in each segment of the figure. 

The training overhead increases quasi-linearly in each segment, but zigzags across different segments. The reason is that \sys's training overhead increases with the number of data points in the experience bucket, and the bucket used for training (i.e., one out of $K^2$) depends on the previous and current protocol. Within one segment, since \sys chooses the best-performing protocol dominantly after convergence, a certain bucket is selected dominantly and accumulates training data. The drops of overhead in the figure is caused by \sys's exploration of other protocol, resulting in other smaller buckets being retrained. On the contrary, the inference overhead is always constant with the number of epochs.

We also measured the duration of epochs in this experiment: $0.88$s in minimum and $1.31$s in average. During the entire $4$-hour deployment, compared to the duration of epochs, even the maximum training and inference overhead is negligible, due to the lightweight nature of random forests. More importantly, the learning agent and validator are two parallel processes. Thus, the overhead of learning does not adversely affect the node's 
throughput, as long as there are some spare CPU cycles devoted to the learning agent. When \sys is deployed for a longer run, techniques such as periodic resampling and limiting the size of the experience bucket can be utilized~\cite{bao} to control the overhead of learning.

%% file: sec_related.tex
\section{Related Work} \label{sec:related}

The BFT consensus problem has been studied extensively, and surveying it is beyond the scope of this work; readers may find introductory material in textbooks, various surveys, and measurement studies, e.g., \cite{attiya2004distributed,malkhi2019concurrency, correia2011byzantine, platania2016choosing, cachin2017blockchain, distler2021byzantine, zhang2022reaching, wang2022bft,singh2008bft}. As an example, BFTSim~\cite{singh2008bft} provides a simulation environment for BFT protocols, which has been used to evaluate a set of protocols under multiple scenarios.
The specific set of BFT protocols which are incorporated in \sys \cite{castro2002practical, kapitza2012cheapbft, amir2011prime, kotla2007zyzzyva, gueta2019sbft, malkhi2023hotstuff} belong to a category of solutions for a setting called partial synchrony which has been introduced in \cite{dwork1988consensus}.

\sys employs a mechanism similar to \textit{reconfiguration} which allows the system to switch the parameters of a BFT algorithm by agreement at certain positions within a sequence of decisions. The reconfiguration problem has been formalized in several works, both in the benign fault settings and the Byzantine setting, e.g., \cite{hunt2010zookeeper,burrows2006chubby,lamport2001paxos,lamport2009vertical,bessani2014state,guerraoui2010next,aublin2015next}. 

To our knowledge, the first work to propose switching between a set of BFT protocols in real-time to adapt to dynamic conditions is \abs ~\cite{aublin2015next}.
It was followed by \adapt \cite{bahsoun2015making}, which further enhanced the approach by training a supervised learning model to govern switching. \sys 
embraces the idea of a multi-protocol BFT engine while considerably enhancing its practicality via a decentralized reinforcement learning engine that provides significant operational benefits and by introducing deep performance metrics, capturing faults, and being hardware agnostic. 

At the core of \sys, learning in untrusted distributed settings is made resilient against intentional data pollution via robust aggregation and agreement. Similar techniques have been explored in robust distributed learning studies \cite{allouah2023robust, yin2018byzantine, karimireddy2021learning, farhadkhani2022byzantine, guerraoui2018hidden, baruch2019little}.
More generally, harnessing learning to enhance performance has been done successfully in many systems under the umbrella of machine programming~\cite{pillars}: indexing~\cite{ml_index}, query optimization~\cite{bao, balsa, lero}, database tuning~\cite{ml_tuning}, software analysis~\cite{controlflag}, scheduling~\cite{decima}, concurrency control~\cite{polyjuice}, and transaction management in blockchains~\cite{wu2022adachain}.

%% file: sec_conc.tex
\section{Conclusion}

Existing BFT protocols lack flexibility and adaptability, leading to suboptimal performance in various scenarios. In this paper, we propose a practical reinforcement learning-based BFT system called \sys, which dynamically selects the top-performing BFT protocols in real-time. Our extensive evaluation demonstrates that \sys significantly outperforms existing solutions under various conditions, including dynamic environments and adversarial attacks.

%% file: seca_background.tex
\section{Background on Selected BFT Protocols} \label{sec:appendix_background}


In this section, we provide background on BFT protocols that are experimentally studied in Section~\ref{sec:landscape}.

\begin{figure}[t]
\centering
\includegraphics[width=0.9\linewidth]{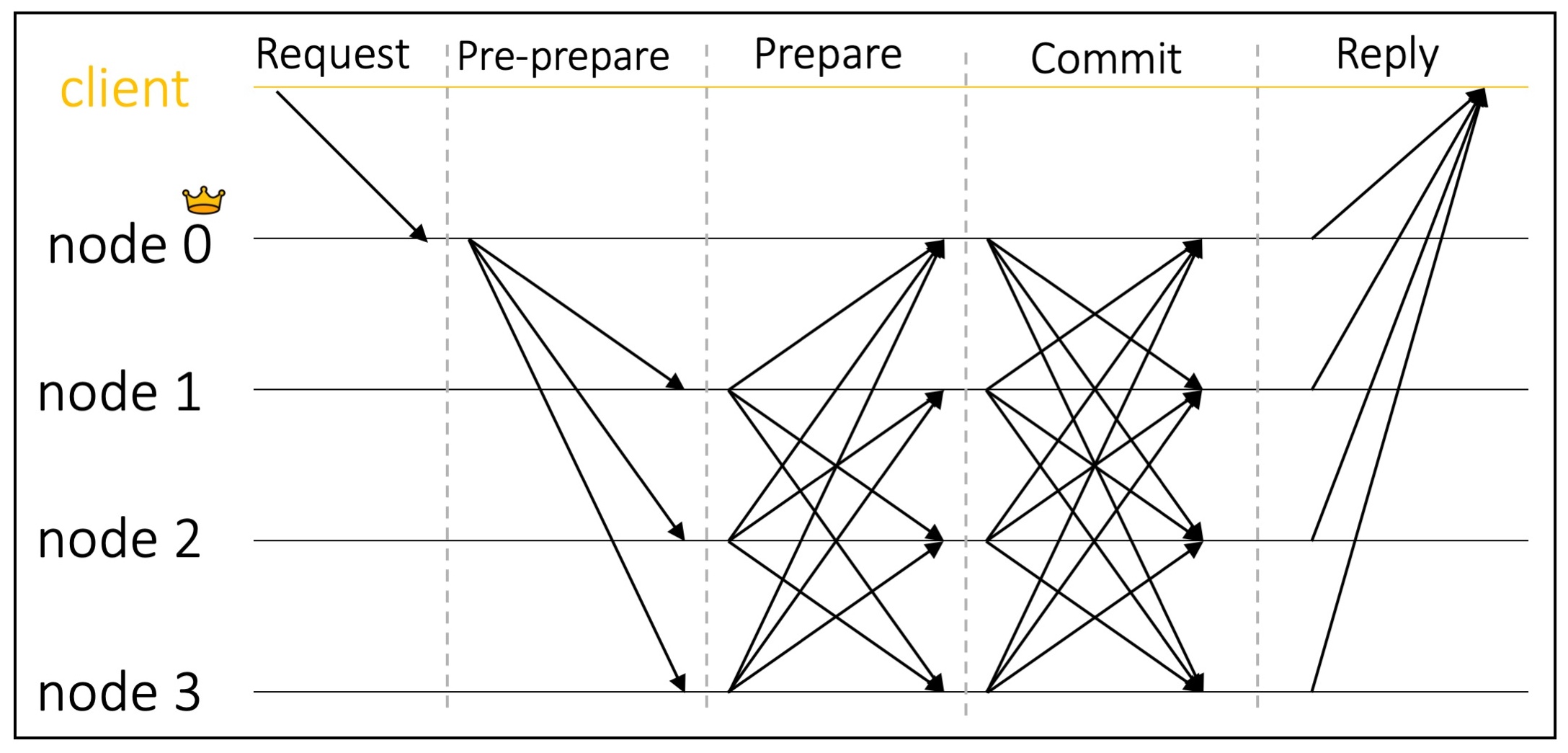}
\vspace{-1em}
\caption{PBFT protocol}
\vspace{1em}
\label{fig:pbft}
\end{figure}

\begin{figure}[t]
\centering
\includegraphics[width=0.9\linewidth]{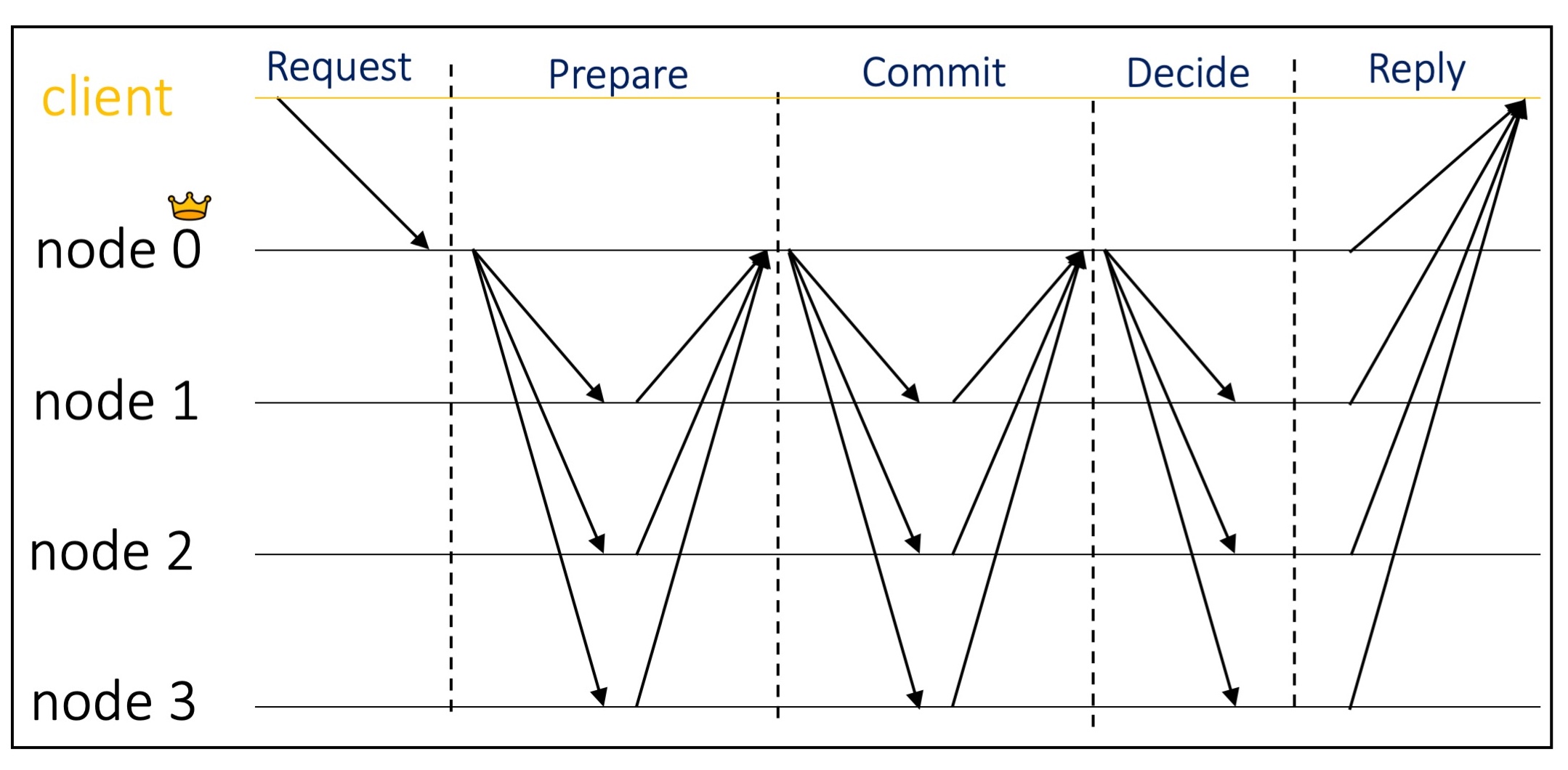}
\vspace{-1em}
\caption{HotStuff-2 protocol}
\vspace{1em}
\label{fig:hotstuff}
\end{figure}

\noindent {\bf PBFT \cite{castro1999practical,castro2002practical}.}
PBFT (Figure~\ref{fig:pbft}) is a leader-based protocol that operates in a 
succession of configurations called {\em views} \cite{el1985efficient,el1985availability}.
Each view is coordinated by a {\em stable} leader (primary).
PBFT consists of \one, \two, and \three phases.
The \one phase assigns an order to the request,
the \two phase guarantees the uniqueness of the assigned order, and
the \three phase guarantees that the next leader after view-change can safely assign the order.

During a normal case execution of PBFT,
clients send their signed \req messages to the leader.
In the \one phase, the leader 
assigns a sequence number to the request to determine the execution order of the request
and multicasts a \one message to all backups.
Upon receiving a valid \one message from the leader,
each backup node multicasts a \two message to all nodes
and waits for \two messages from $2f$ different nodes (including the node itself) that match the \one message.
The goal of the {\sf prepare} phase is to guarantee safety within the view, i.e.,
$2f$ nodes received matching \one messages from the leader node and agree with the order of the request.

Each node then multicasts a \three message to all nodes.
Once a node receives $2f+1$ valid \three messages from different nodes, including itself,
that match the \one message, it commits the request.
The goal of the \three phase is to ensure safety across views, i.e., 
the request has been voted on a majority of non-faulty nodes and can be recovered after (leader) failures.
The second and third phases of PBFT follow the {\em clique} topology, i.e., have $O(n^2)$ message complexity.
If the node has executed all requests with lower sequence numbers, it executes the request and
sends a \reply to the client.
The client waits for $f+1$ matching results from different nodes.

\noindent
{\bf HotStuff-2 \cite{malkhi2023hotstuff}.}
HotStuff-2 (Figure~\ref{fig:hotstuff}) HotStuff-2 is a leader-based BFT protocol with two main properties.
First, it provides linear communication complexity (rather than quadratic as in PBFT).
Specifically, each all-to-all communication phase of PBFT is replaced with two linear phases in HotStuff;
one from the replicas to the leader and one from the leader to the replicas.
Second, HotStuff-2 uses the leader rotation technique, where the leader is replaced
after every single proposal in a predetermined manner (round-robin).
This is in contrast to most existing protocols that rely on a stable leader,
and the leader is changed only when it is suspected to be faulty.
Compared to the original HotStuff \cite{yin2019hotstuff}, HotStuff-2 reduces one phase of communication, improving the overall performance of the protocol.
Chaining (also proposed in \cite{yin2019hotstuff}) is an optimization technique that is applicable to both HotStuff and HotStuff-2. The chaining optimization involves executing some phases which are deemed identical in a pipelined manner in order to reduce the latency of request processing.

\begin{figure}[t]
\centering
\includegraphics[width=0.5\linewidth]{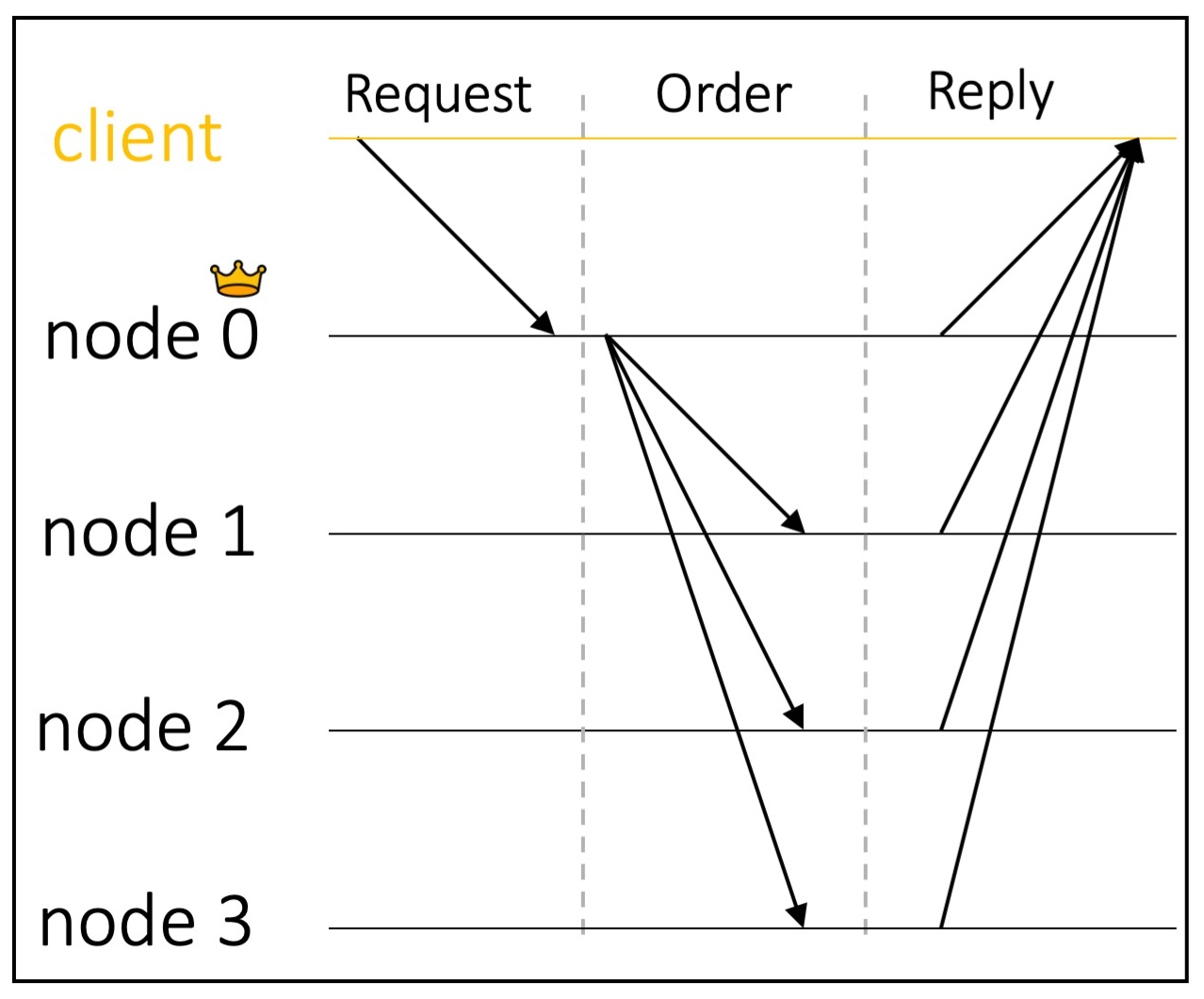}
\vspace{-1em}
\caption{Zyzzyva protocol (fast path)}
\vspace{1em}
\label{fig:zyzzyva}
\end{figure}

\begin{figure}[t]
\centering
\includegraphics[width=0.75\linewidth]{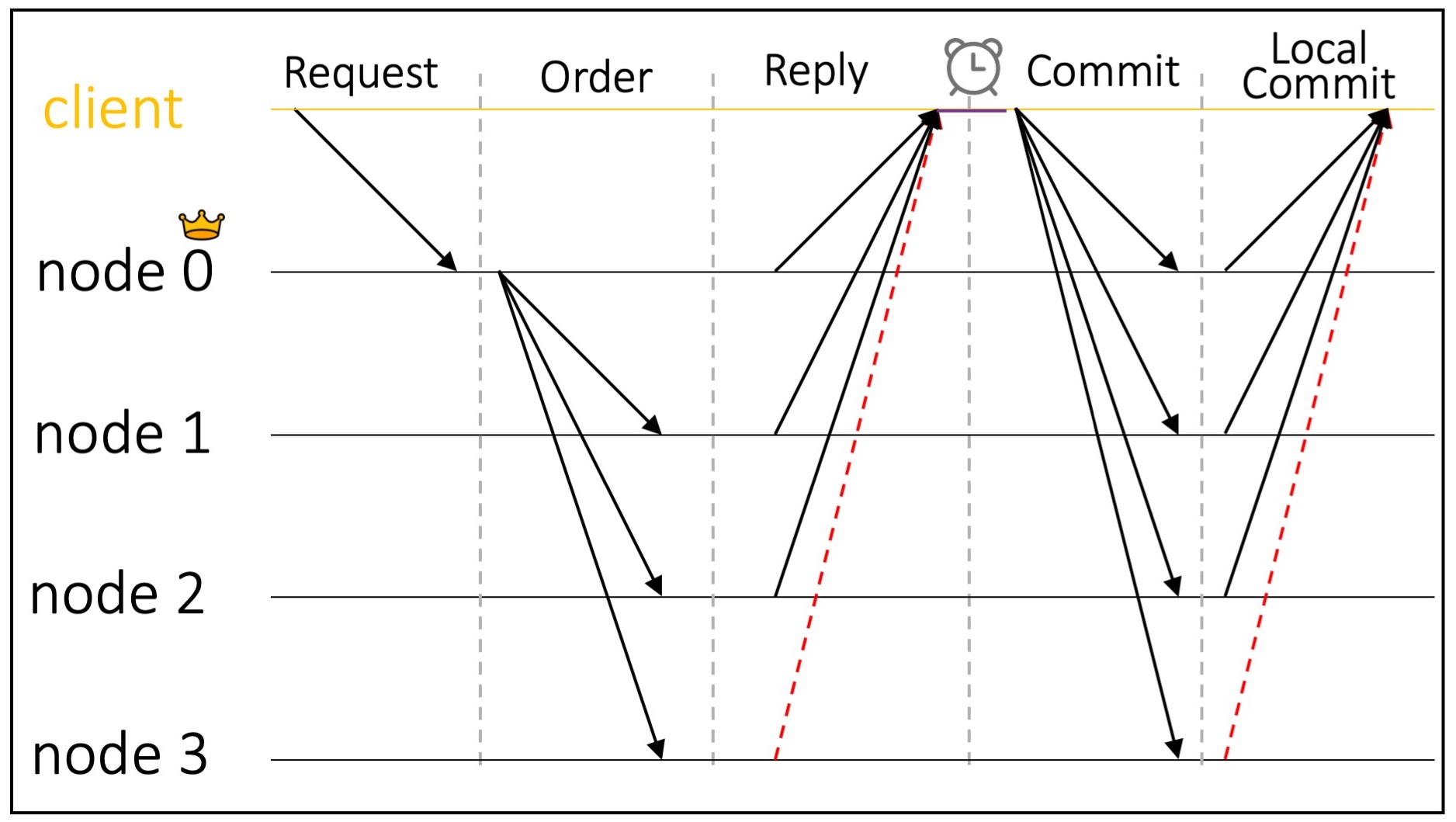}
\vspace{-1em}
\caption{Zyzzyva protocol (slow path)}
\vspace{1em}
\label{fig:zyzzyva-s}
\end{figure}

\noindent
{\bf Zyzzyva \cite{kotla2007zyzzyva}.}
Zyzzyva
(Figure~\ref{fig:zyzzyva}) optimistically assumes that the leader and all backups are non-faulty.
Hence, upon receiving an \order message from the primary node which includes the request, nodes speculatively execute requests without running any agreement and send \reply messages to the client.
The client waits for $3f+1$ matching replies to accept the results.
If the client timer is expired and the client has received matching replies from between $2f+1$ and $3f$ nodes,
as presented in Figure~\ref{fig:zyzzyva-s},
two more linear rounds of communication are needed to ensure that at least $2f+1$ nodes
have committed the request.

\begin{figure}[t]
\centering
\includegraphics[width=0.7\linewidth]{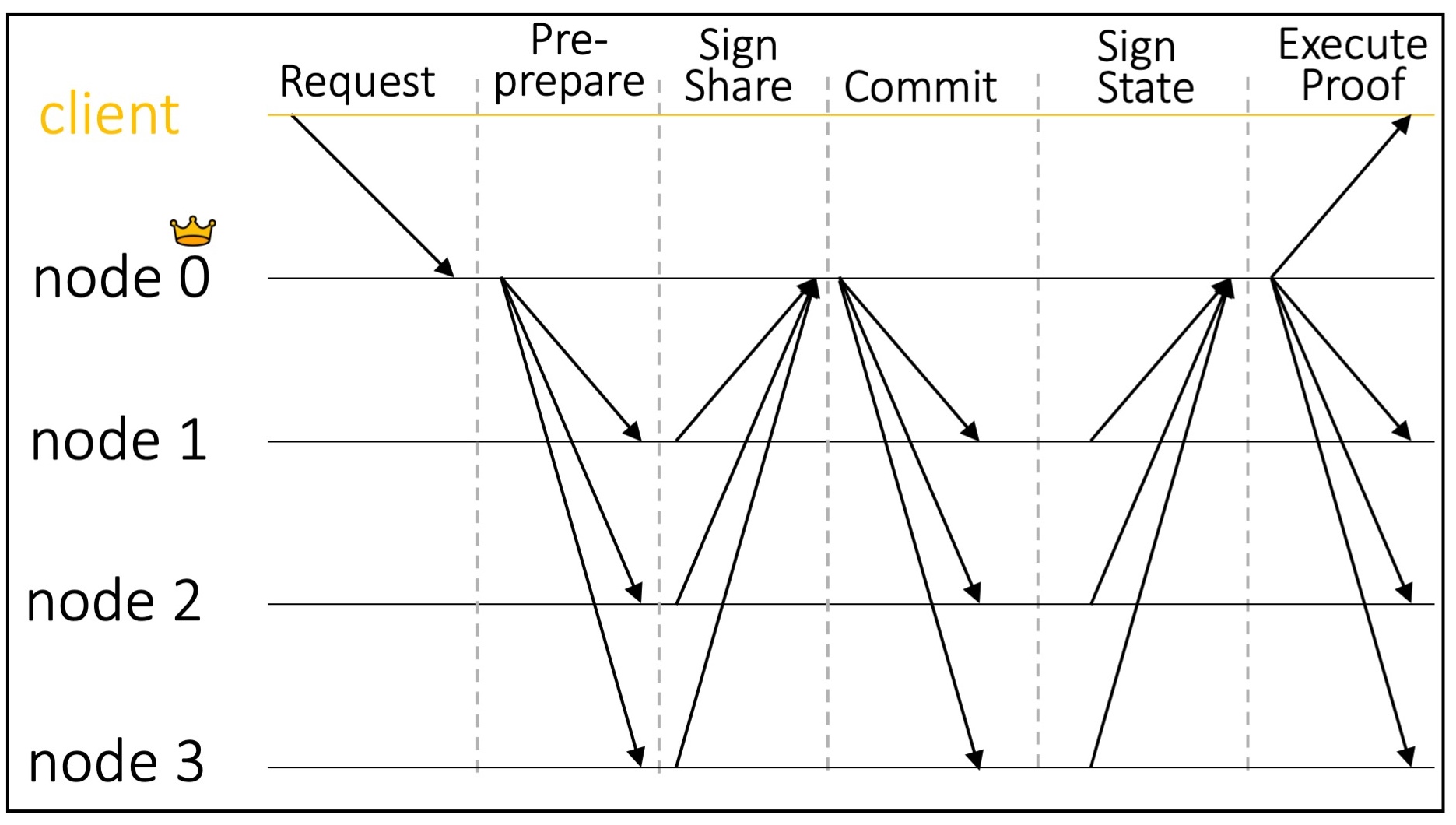}
\vspace{-1em}
\caption{SBFT protocol (fast path)}
\vspace{1em}
\label{fig:sbft}
\end{figure}

\begin{figure}[t]
\centering
\includegraphics[width=0.9\linewidth]{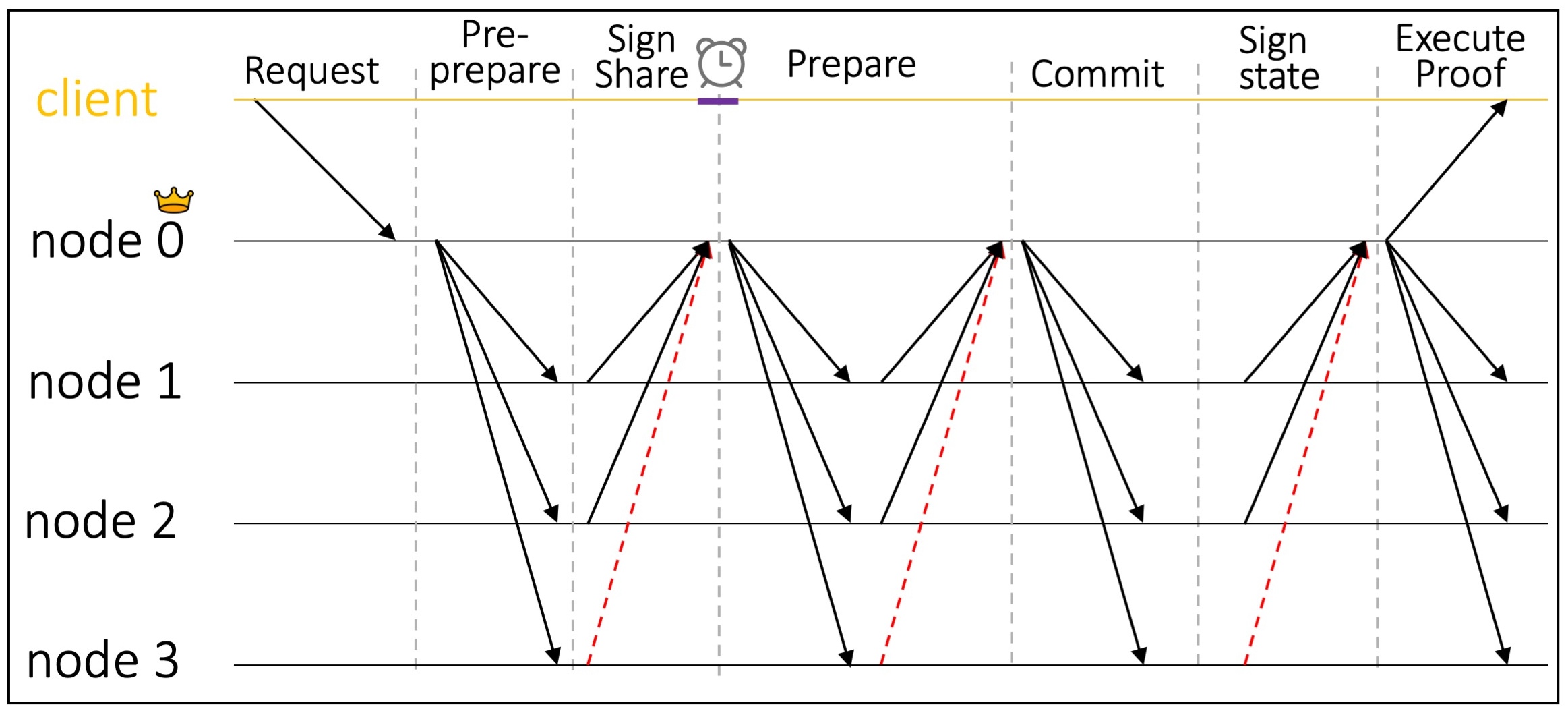}
\vspace{-1em}
\caption{SBFT protocol (slow path)}
\vspace{1em}
\label{fig:sbft-s}
\end{figure}

\noindent
{\bf SBFT \cite{gueta2019sbft}.}
SBFT\footnote{SBFT tolerates both crash and Byzantine failure
($n=3f+2c+1$ where $c$ is the number of crashed nodes). Since the focus of this paper is on Byzantine failures, we consider a variation of SBFT where $c=0$.}
presents a linear optimistic fast path (Figure~\ref{fig:sbft}), assuming all nodes are non-faulty.
In SBFT, upon receiving a \one message from the primary node, all nodes send \share messages to the commit collector (i.e., the primary in our figure). If the commit collector is able to collect $3f+1$ \share messages, it puts them together to generate a \three message and broadcasts it to the backups.
Otherwise (the commit collector does not receive messages from {\em all} backups in the \share phase and its timer is expired),
SBFT switches to its slow path (Figure~\ref{fig:sbft-s}) and requires two more linear rounds of communication (\two phase).
The dual-path nature of SBFT requires nodes to sign each message with two schemes (i.e., $2f+1$ and $3f+1$).
To send replies to the client,
a single (execution collector) node receives replies from all nodes and sends a single (threshold) signed reply message.

\begin{figure}[t]
\centering
\includegraphics[width=0.7\linewidth]{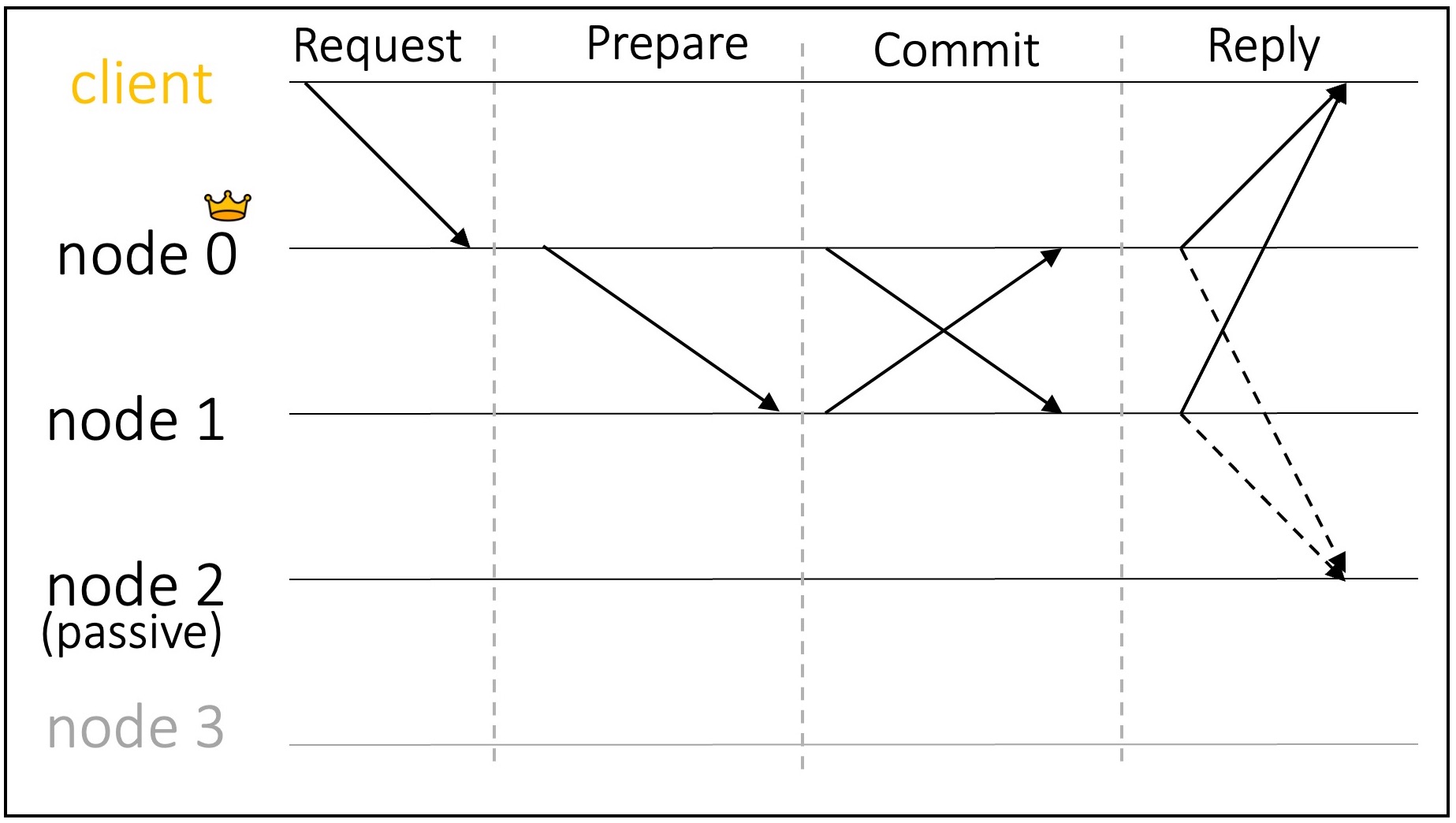}
\vspace{-1em}
\caption{CheapBFT protocol}
\vspace{1em}
\label{fig:cheapbft}
\end{figure}

\noindent
{\bf CheapBFT \cite{kapitza2012cheapbft}.}
CheapBFT (Figure~\ref{fig:cheapbft}) differs from PBFT in two major ways.
First, it puts $f$ nodes as passive nodes by optimistically assuming all nodes participating in the quorum (i.e., called \textit{active} nodes) are honest. Second, it relies on trusted hardware to prevent equivocation. CheapBFT reduces the required quorum of active nodes to $f+1$ (the total number of nodes becomes $2f+1$) in its normal case execution, and reduces the number of phases from three to two as compared with PBFT (i.e., no \one phase). If an active node becomes faulty, it is replaced with a passive node.

\begin{figure}[t]
\centering
\includegraphics[width=\linewidth]{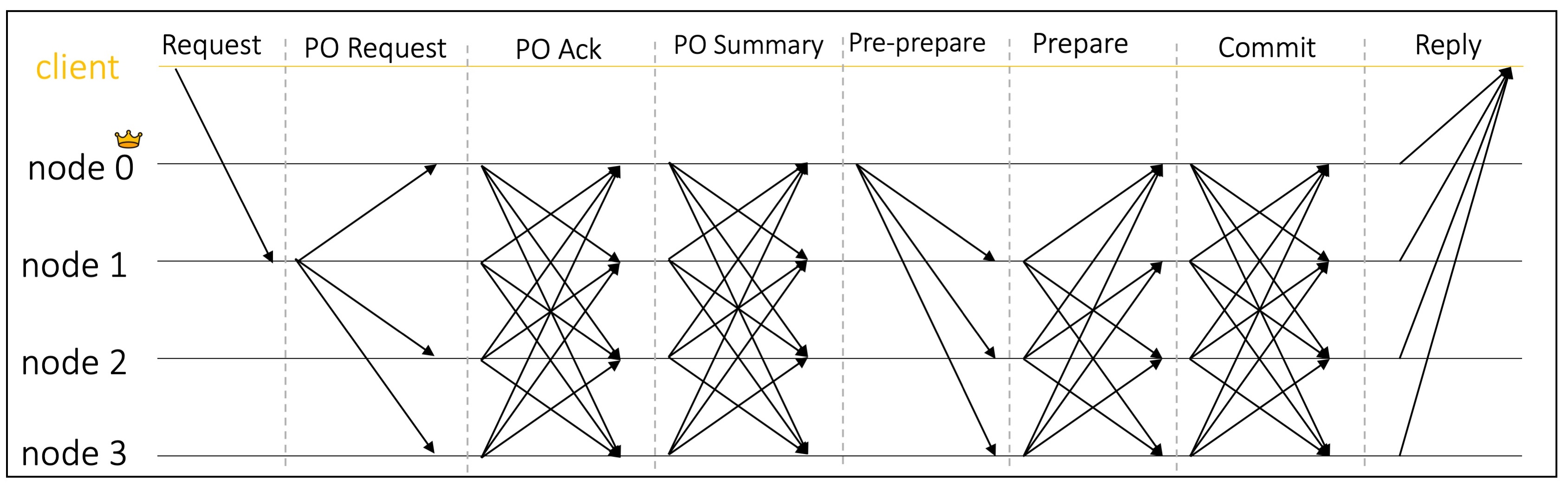}
\vspace{-1em}
\caption{Prime protocol}
\vspace{1em}
\label{fig:prime}
\end{figure}

\noindent
{\bf Prime \cite{amir2011prime}.}
Prime (Figure~\ref{fig:prime}) is a robust protocol by incorporating a pre-ordering stage.
In the pre-ordering stage, nodes exchange the requests they receive from clients and
periodically share a vector of all received requests, expecting the leader to globally order requests following those vectors.
In this way, nodes can also monitor the leader to order requests in a fair manner.
Prime also uses timers to periodically check performance of the protocol and replace the primary node if it does not provide acceptable performance (e.g., under slowness attack) corresponding to the current network condition.

%% file: seca_switch.tex
\section{Switching BFT Protocols} \label{sec:appendix_switching}

In this section, we describe how \sys performs the actual switching from BFT protocol $t$ to $t+1$. 

Similar to \abs, \sys operates in epochs. Each epoch in \sys can be considered as a \textit{Backup} instance in \abs, which is marked by the completion of $k$ requests, where $k$ is a predefined constant hyper-parameter. For each epoch, an existing BFT protocol in our action space (i.e., PBFT, Zyzzyva, CheapBFT, Prime, SBFT, and HotStuff-2) is selected by the learning agent and remains unchanged throughout the epoch.

The \textit{Backup} instance in \abs is a thin wrapper that can be put around an existing BFT protocol. \textit{Backup} works as follows: it is inactive until it receives a request containing a valid \init, which is an unforgeable history generated by the preceding BFT instance. At this point, \textit{Backup} sets its state by executing the requests in the \init it received. Then, it simply executes the first $k$ requests committed by the underlying BFT protocol, aborts all subsequent requests, and returns the signed sequence of executed requests as the \init for the next BFT instance. A client can switch from a \textit{Backup} instance $t$ to instance $t+1$ as soon as it receives $f{+}1$ signed messages from different replicas, containing an identical \init. The client uses the \init to invoke instance $t{+}1$. Once instance $t$ aborts some requests and switches to $t+1$, it cannot commit any subsequently invoked requests.  


\noindent \textbf{Optimizations of Abstract.}
\abs also implements the following optimization for the switching procedure above: align switching from a \textit{Backup} instance with a checkpoint in order to minimize the size of \init, 
which is transferred through clients. In \abs, BFT instances $t$ and $t+1$ might be run on two different clusters, and not every instance is a \textit{Backup} instance whose progress condition is $k$ committed requests. For example, an instance could be \textit{Quorum} whose progress condition is $3f+1$ matching history, or \textit{Chain} whose progress condition is one correct reply, both of which require an honest client to validate the progress condition and initiate the switching. On the contrary,  in \sys, all BFT instances are run on the same cluster of machines, and every BFT instance is a \textit{Backup} instance. Thus, the optimization can be taken one step further inside \sys. First, instead of relying on the client, each replica has enough information to decide whether to switch or not. Once executing $k$ requests, each replica multicasts an \init (i.e., checkpoint) to all other replicas. Second, an honest replica does not need to be blocked and wait for $f+1$ signed matching \init and execute it before starting the new epoch, since the \init is already reflected in its local service state. Once $k$ requests are committed in epoch $t$ and the BFT protocol for $t+1$ is derived by the learning agent, epoch $t+1$ is invoked. In other words, the switching can be performed asynchronously inside \sys with low overhead. 

\input{tbl_landscape}

\noindent \textbf{Speculative Backup instances.}
\textit{Backup} can be easily implemented over an existing BFT protocol, except for speculative protocols where clients are the commit collectors, e.g., Zyzzyva. This is because such protocols rely on clients to decide if a request has been committed, while the replicas might not have enough information to tell. Specifically, in Zyzzyva, if a request commits in the slow path where the client only receives $2f+1$ replies with matching history (instead of $3f+1$ replies in its fast path) and multicasts a commit certificate to all replicas, the replicas can deterministically commit the request as well. However, if a request commits in the fast path, the client directly completes the request without notifying the replicas. Thus, the replicas can not decide whether the current epoch is finished.

To address this issue, when running speculative protocols in epoch $t$, \sys enforces the $k$-th request (i.e., the last request in $t$) to be a special {\footnotesize \textsf{NOOP}} request and to be committed in the slow path. On each replica, the view-change timer for this request is only removed when it can be considered as committed. Specifically, for the $k$-th request, the leader acts as a dummy client and multicasts a commit certificate even if $3f+1$ replies with matching history are received. Upon receiving the commit certificate or $f+1$ signed matching \init for epoch $t+1$, replicas can safely consider the $k$-th request of epoch $t$ as completed. Note that a malicious dummy client might refuse to multicast the commit certificate. In this case, view-change is triggered and a new leader will be elected, which will act as a new dummy client for this request. Then, in the {\footnotesize \textsf{NEW-VIEW}} message, for this $k$-th request, the new leader always inserts a {\footnotesize \textsf{ORDER-REQ}} message where the original {\footnotesize \textsf{NOOP}} request is replaced by the {\footnotesize \textsf{NOOP}} request proposed by its own dummy client. For this special request, no safety issues exist since its content is not part of the service state. The mechanisms above guarantee that each honest replica will successfully commit this request (i.e., liveness), and thus concluding the current epoch.

\noindent \textbf{Correctness of switching.}
Below, we discuss the safety and liveness of switching between epochs.

\noindent \underline{\textit{Safety.}} \abs's idempotency theorem~\cite{aublin2015next} specifies that if individual BFT instances are correct, irrespectively of each other, then the system composed through switching is also correct. Since each BFT instance in \sys is an existing protocol whose safety has been previously proven (instead of our newly invented protocol), the composed system is safe.

\noindent \underline{\textit{Liveness.}}  \abs guarantees liveness if a request is not aborted by all instances, which can be made simple by reusing an existing BFT protocol as one of the instances. Apparently, \sys satisfies this requirement too.

Moreover, to ensure liveness, \abs exponentially increases the parameter $k$ with every new instance of \textit{Backup} by default. This is to prevent a corner case with very slow clients: $k$ requests committed by a single \textit{Backup} instance $i$ might all be invoked by the same, fast client, while a slow client can then get its requests aborted by $i$. The same can happen with a subsequent \textit{Backup} instance, and so forth. By exponentially increasing $k$, this liveness issue is resolved since no realistic load increases faster than exponentially. Alternatively, by having the replicas across different \textit{Backup} instances share a client input buffer, such an issue can also be prevented without exponentially increasing $k$. \sys takes the latter approach, since all epochs run on the same cluster, so the client input buffer is naturally shared across epochs.

%% file: tbl_landscape.tex
\begin{table*}[t]
\caption{The throughput of each studied protocol under different conditions. Each row characterizes a different condition, where the best throughput is highlighted in {\color{blue}\textbf{blue}}.}
\centering
\vspace{0.5em}
\scriptsize
\begin{tabular}{ccccc|rrrrrr}
\toprule
\multicolumn{5}{c|}{Condition Parameters} & \multicolumn{6}{c}{Throughput (tps)} \\
f & \# of clients & \# of absentees & request size & proposal slowness & PBFT & Zyzzyva & CheapBFT & Prime & SBFT & HotStuff-2 \\ \midrule

1 & 50 & 0  &4KB & 0ms & 9133 & {\color{blue}\textbf{13664}} & 11822 & 4601 & 11067 & 6882 \\ 
4 & 100 & 0  &4KB & 0ms  & 4316 & {\color{blue}\textbf{10699}} & 7966 & 4239 & 6414 & 7124  \\ 
4 & 100 & 0  &100KB & 0ms & 4261 & 6513 & {\color{blue}\textbf{7353}} & 4177 & 6518 & 6779  \\ 
4 & 100 & 4  &4KB & 0ms & 5386 & 1929 & {\color{blue}\textbf{10011}} & 4440 & 5347 & 8848  \\ 
4 & 100 & 0  &0KB & 20ms & 2435 & 2424 & 2433 & 4265 & 2432 & {\color{blue}\textbf{6201}}  \\ 
4 & 100 & 0  &1KB & 20ms & 2435 & 2424 & 2432 & 4211 & 2433 & {\color{blue}\textbf{6099}}  \\ 
4 & 100 & 0  &0KB & 100ms & 497 & 498 & 497 & {\color{blue}\textbf{4257}} & 497 & 3641  \\ 
1 & 50 & 0  &0KB & 20ms & 989 & 988 & 989 & {\color{blue}\textbf{4527}} & 989 & 2640  \\ 
\bottomrule
\end{tabular}
\label{tbl:landscape_full}
\end{table*}

%% file: seca_collect.tex
\section{Learning Coordination} \label{sec:appendix_collect}

In this section, we present \sys's learning coordination protocol and discuss the correctness of \sys. 

\subsection{Detailed Description of The Protocol} \label{sec:appendix_protocol}

Algorithm~\ref{alg:collect} presents the detailed learning coordination protocol, where VBC (i.e., any ``blackbox'' validated Byzantine consensus primitive) is implemented using PBFT.
As shown in lines 1-7, after executing $w$ requests (a hyper-parameter) in epoch $t$, each node $i$ collects local performance indicators $p_i^{t-1}$ measured during epoch $t-1$, featurizes the next state $f_i^{t+1}$, and broadcasts both metrics inside a \rpt message.
Each agent $i$ collects the received valid \rpt messages where $p_j^{t-1}$ and $f_j^{t+1}$ are non-null, into a local set $reportQC_i^t$.
Upon collecting $f+1$ \rpt messages from different agents, the agent triggers the coordination protocol for epoch $t$ and starts a view-change timer $\tau_{c,1}$ to track the progress (lines 8-11).

Each time an agent $l$ becomes the leader, as shown in lines 12-15 (the view-change leader is omitted), it starts an additional timer $\tau_{c,2}$ for collecting $2f+1$ reports.
Once the size of  $reportQC_l^t$ reaches $2f+1$ or $\tau_{c,2}$ expires, the leader multicasts 
 $\langle \PRO, v_c, n_c, t, d \rangle_{\sigma_l}, reportQC_l^t\rangle$ message
to all agents, where $v_c$ is the view in the coordination protocol, $n_c$ is the sequence number in the coordination protocol, $t$ is the epoch id, and $d$ is the digest of $reportQC_l^t$.

Once a coordinator agent $i$ receives a valid \pro with coordination sequence $n_c$ that it has not voted for, it accepts that proposal and multicasts a \prp message if:
(1) the size of corresponding $reportQC_l^t$ is at least $f+1$,
(2) $t$ is not committed and
(3) $n_c-1$ is committed (lines 16-18).

Then, similar to PBFT, after the \prp and \cmt phases, the predicate {\sf c-committed}$(v_c, n_c, t, reportQC_l^t)$ becomes true if $2f+1$ matching \cmt messages from different agents have been received. Once committed, agent $i$ checks the size of $reportQC_l^t$. If smaller than $2f+1$, the learning algorithm (i.e., retraining and inference based on predictive models) is not invoked, and the agent utilizes the same protocol for epoch $t+1$ as in the current epoch $t$. It also initiates a view-change on the current BFT protocol used by its validator and multicasts a \vc message to all other agents. Here, $s_c$ stands for the last stable checkpoint known to $i$ in the coordination protocol, $C_c$ is a set of \cpt messages proving the correctness of $s_c$, and $P_c$ is a set containing a set $P_{reportQC}$ for each $reportQC$ that prepared at $i$ with a coordination sequence number higher than $s_c$. Each  $P_{reportQC}$ contains the \pro and \prp messages (lines 23-26). Otherwise, if the size is $2f+1$, for reward and each state dimension, the median value is taken as the global reward $p^{t-1}$ and state $f^{t+1}$. The state-action-reward triplet for epoch $t-1$ is then added to the experience buffer, the predictive models are re-trained, and a promising BFT protocol is inferred for epoch $t+1$ based on state $f^{t+1}$ (lines 27-31). Finally, regardless of the size of $reportQC_l^t$, a \cpt message is multicast to all coordination agents.

\input{alg_data_collection}

Upon timer $\tau_{c,1}$ expires, each coordination agent multicasts a \vc message to all agents. Once receiving $2f+1$ \vc messages from different agents, the new leader $l'$ begins the \nv phase similar to PBFT. At the end of this phase, $l'$ checks all $P_{reportQC}$ of $P_c$ components collected in the view-change quorum. If none of them includes a $reportQC$ for epoch $t$, $l'$ will multicast a \pro message to all agents which carries the $reportQC_{l'}^t$ it has collected.

We argue that the leader agent could also deliberately delay \pro messages. However, unlike the active BFT instance in each epoch that has tens of thousands of requests to commit, this coordination protocol runs only once every epoch. Thus, the performance impact of such malicious behavior is negligible.

\subsection{Proof Sketch} \label{sec:appendix_proof}

Below, we formalize the safety, liveness, and robustness guarantees of \sys, followed by the proof sketch.

\noindent \textbf{Safety.}  For each epoch $t$, each honest learning agent will agree on the same feature and the same reward if its learning algorithm is invoked.

\begin{proof}
    We prove by contradiction that only one $reportQC$ can be committed for the same epoch $t$. First, by reduction to PBFT, \pro, \prp, \cmt phases guarantee that for each coordination sequence $n_c$, only one $reportQC$ can be committed. Then, assume two different $reportQC$ for the same epoch $t$ are committed at different $n_c$: $reportQC_j^t$ committed at $n_{c,1}$, and $reportQC_k^t$ at $n_{c,2}$, where $n_{c,1}<n_{c,2}$ and $D(reportQC_j^t) \neq D(reportQC_k^t)$. At least one honest agent $i$ needs to accept the \pro for both $reportQC$. Since the network is asynchronous, we consider two cases. If $n_{c,1}$ is committed before $n_{c,2}$ on $i$, $i$ needs to accept \pro with $reportQC_k^t$ when $t$ is already committed. Otherwise if $n_{c,1}$ is committed at the same time with or after $n_{c,2}$, $i$ needs to accept $reportQC_k^t$  whose coordination sequence is $n_{c,2}$, before $n_{c,2}{-}1$ is committed. Both cases contradict with line 17 of Algorithm~\ref{alg:collect}. Thus, only one $reportQC$ can be committed for the same epoch $t$. Finally, since an honest learning agent takes the median value of the same committed $reportQC$, they agree on the same feature and reward if $reportQC$ includes $2f+1$ reports. 
\end{proof}

Following this safety property, through deterministic training (i.e., using the same random seed on each learning agent), each honest learning agent in \sys derives the same action (i.e., BFT protocol) for the same epoch $t$.

\noindent \textbf{Liveness.} \sys guarantees that every honest agent eventually invokes its learning algorithm in the same epoch. 

\begin{proof}
If fewer than $2f+1$ reports are collected, the learning algorithm will be ineffective (i.e. not invoked) for one epoch, and the BFT protocol chosen for the next epoch will be the same as the current one. We first show agents are able to commit a $reportQC$ of $2f+1$ reports within $f$ consecutive epochs (Part 1). We then show if such $reportQC$ is committed for epoch $t$, every honest agent will invoke its learning algorithm in epoch $t$ (Part 2).

\noindent \underline{Part 1:} In the active BFT protocol of any certain epoch, there can be at most $f$ honest validators placed in-dark by the malicious leader, since otherwise, $f+1$ validators will trigger a view-change to replace the leader in the epoch. The agents on these in-dark validators will not send \rpt messages according to our protocol. In addition, at most $f$ malicious agents can deliberately refuse to send \rpt messages. Thus, each agent is guaranteed to receive at least $f+1$ reports for epoch $t$, which starts a timer $\tau_{c,2}$ on leader agent $l$ for collecting $2f+1$ reports in $reportQC_l^t$ and a timer $\tau_{c,1}$ on each agent for tracking the progress of coordination protocol. If $reportQC_l^t$ is committed, a view-change in the active BFT protocol used by validators will be initiated if it contains less than $2f+1$ reports. Since such coordination is run once every epoch, an honest leader validator will be found within at most $f$ consecutive epochs. Otherwise, if $reportQC_l^t$ is not committed, the leader agent will be replaced while the current epoch waits for a report quorum to be committed.

\noindent \underline{Part 2:} If a $reportQC$ of size $2f+1$ is committed on an agent, it is committed on at least $2f+1$ agents, otherwise \vc is triggered and it will be committed in view $v_c+1$ or above. Thus, at least $f+1$ honest agents will multicast a \cpt message containing the $reportQC$ as the service state of the coordination protocol. Every honest agent, even if it is placed in dark by the malicious leader agent, will receive this stable checkpoint and thus invoke its learning algorithm.

\end{proof}

\noindent \textbf{Robustness.} If honest feature/reward values form a range [$r_l, r_h$], the global feature/reward taken by the learning agent always falls into this range.

\begin{proof}
According to our protocol, honest nodes never report invalid (e.g., zero or null) values to others, even when they are placed in-dark and recover their service state from others.

The learning agent only takes the median value of $reportQC$ of size $2f+1$, where at most $f$ can be arbitrary values reported by malicious replicas, and the remaining $f+1$ is guaranteed to be honest. Now we prove by contradiction: w.l.o.g., assume the median value of $reportQC$ is $r_m$, where $r_m>r_h$. Since there are at least $f+1$ honest values smaller than $r_m$, to make $r_m$ the median, at least $f+1$ dishonest values larger than $r_m$ need to be reported. This contradicts the fact that, at most $f$ can be arbitrary values in $reportQC$.
\end{proof}

%% file: alg_data_collection.tex
\newcommand{\RPT}{\footnotesize \textsf{REPORT}\xspace}
\newcommand{\PRP}{\footnotesize \textsf{C-PREPARE}\xspace}
\newcommand{\CMT}{\footnotesize \textsf{C-COMMIT}\xspace}
\newcommand{\CPT}{\footnotesize \textsf{C-CHECKPOINT}\xspace}
\newcommand{\VC}{\footnotesize \textsf{C-VIEW-CHANGE}\xspace}

\begin{algorithm}[t]
\footnotesize
\caption{Learning coordination}\label{alg:collect}
\begin{algorithmic}[1]

\Statex {\color{teal} $\rhd$ On each agent $i$}
\State \textbf{Upon} execution of $w$ requests in epoch $t$
\State \quad \textbf{if} no state transfer happens during epoch $t-1$ \textbf{then}
\State \qquad Record performance $p_i^{t-1}$
\State \quad \textbf{if} no state transfer happens during epoch $t$ \textbf{then}
\State \qquad Extract features $f_i^{t+1}$ from executed requests in window $w$
\State \quad \textbf{if} $p_i^{t-1} \neq null \land f_i^{t+1} \neq null $ \textbf{then}
\State \qquad Multicast  $\langle \RPT, t, i, p_i^{t-1}, f_i^{t+1}\rangle_{\sigma_i}$ to all agents
\State \textbf{Upon receiving} valid \RPT message $m$
\State \quad Add $m$ to $reportQC_i^t$ 
\State \quad \textbf{if} $reportQC_i^t.size=f+1$ \textbf{then}
\State \qquad Start timer $\tau_{c,1}$
\Statex {\color{teal} $\rhd$ On the leader agent $l$}
\State \textbf{Upon} $reportQC_l^t.size$ reaching $f+1$
\State \quad Start timer $\tau_{c,2}$ \Comment{$\tau_{c,2}<\tau_{c,1}$}
\State \textbf{Upon} timer $\tau_{c,2}$ timeouts $\lor$ $reportQC_l^t.size=2f+1$ 
\State \quad Multicast $\langle \PRO, v_c, n_c, t, d \rangle_{\sigma_l}, reportQC_l^t\rangle$ to all agents
\Statex {\color{teal} $\rhd$ On each agent $i$}
\State  \textbf{Upon receiving} a valid \PRO from the leader
\State \quad \textbf{if} $reportQC_l^t.size \geq f+1 \land t$ is not committed $\land n_c-1$ is committed \textbf{then}
\State \qquad Multicast $\langle \PRP, v_c, n_c, t, d, i\rangle_{\sigma_i}$ to all agents
\State \textbf{Upon receiving} valid matching \PRP from $2f+1$ different agents
\State \quad Multicast $\langle \CMT, v_c, n_c, t, d, i\rangle_{\sigma_i}$ to all agents
\State \textbf{Upon receiving} valid matching \CMT from $2f+1$ different agents
\State \quad Commit ($v_c, n_c, t, reportQC_l^t$)
\State \quad \textbf{if} $reportQC_l^t.size<2f+1$ \textbf{then}
\State \qquad $protocol^{t+1}$ $\gets$ $protocol^t$
\State \qquad Initiate view change on validator $i$ in $protocol^t$
\State \qquad Multicast $\langle \VC, v_c+1, s_c, C_c, P_c, i\rangle_{\sigma_i}$ to all agents
\State \quad \textbf{else} 
\State \qquad $p^{t-1} \gets \text{median}\{m.p_j^{t-1} | m \in reportQC_l^t\}$
\State \qquad $f^{t+1} \gets \text{median}\{m.f_j^{t+1} | m \in reportQC_l^t\}$
\State \qquad Add ($f^{t-1}$, $protocol^{t-1}$, $p^{t-1}$) to experience buffer
\State \qquad $protocol^{t+1}$ $\gets$ \textproc{BestPredictedProtocol}($f^{t+1}$)
\State \quad Multicast $\langle \CPT, n_c, t, i, reportQC_l^n\rangle_{\sigma_i}$ to all agents
\State \textbf{Upon} timer $\tau_{c,1}$ timeouts
\State \quad Multicast $\langle \VC, v_c+1, s_c, C_c, P_c, i\rangle_{\sigma_i}$ to all agents
\Statex {\color{teal} $\rhd$ On the leader agent $l'$ of view $v_c+1$}
\State \textbf{if} no set in the $P_c$ components has $reportQC_l^t$ \textbf{then}
\State \quad Multicast $\langle \PRO, v_c+1, n_c, t, d \rangle_{\sigma_{l'}}, reportQC_{l'}^t\rangle$ to all agents

\Comment{The remaing view change routine are omitted here}

\end{algorithmic}
\end{algorithm}

%% file: seca_results.tex
\section{Additional Experiment Results}

\subsection{Comprehensive Performance Comparison} \label{sec:appendix_table}

Table~\ref{tbl:landscape_full} presents the comprehensive performance comparison results, which were previously summarized by Table~\ref{tbl:landscape} in Section~\ref{sec:landscape}. Under each different condition (i.e., row), we list the throughput of each protocol in terms of transactions per second. For a fair comparison, the common internal parameters of all six protocols are configured with the same value: we set the batch size to be $10$ and the view-change timer to be $100$ms. The protocol-specific internal parameters (e.g., leader rotation interval of HotStuff-2, the aggregation delay for global ordering in Prime, the timer that distinguishes fast path vs. slow path in Zyzzyva) are configured with a reasonable value such that the protocol has good performance on all conditions.

\subsection{Randomized Sampling} \label{sec:appendix_randomized}
\begin{figure}[t]
\centering
\includegraphics[width=0.85\linewidth]{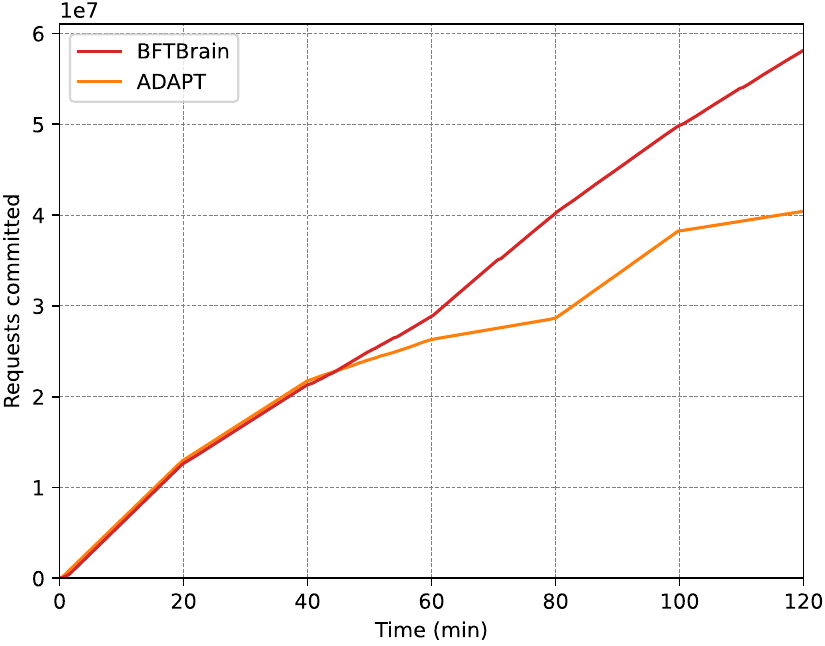}
\vspace{-1em}
\caption{Adaptivity of \sys compared to \adapt under randomly sampled conditions.}
\vspace{1em}
\label{fig:randomized_cdf}
\end{figure}
Compared to the ``cycle back conditions'' experiment in Section~\ref{sec:exp_changing}, we created a benchmark that has more variations in the state space and introduced the variations more frequently. Specifically, each dimension in category State 1 and 2 (except $F1$) follows a certain normal distribution independently. We vary each dimension every 1s by randomly sampling from its distribution, and shift the mean and variance of such distribution every 20 minutes. During the first hour of the experiment, all validators ($n=13$) are responsive, whereas during the second hour, $f$ validators are non-responsive. The entire experiment lasts 2 hours. Compared to other dimensions, we varied $F1$ at a lower frequency, since in real deployment scenarios, client workloads vary more frequently than faults in the system. The exact distribution we used and their shifting pattern can be found here\footnote{\url{https://github.com/JeffersonQin/BFTBrain/tree/master/exp/randomize}}.

\begin{figure}[t]
\centering
\includegraphics[width=0.9\linewidth]{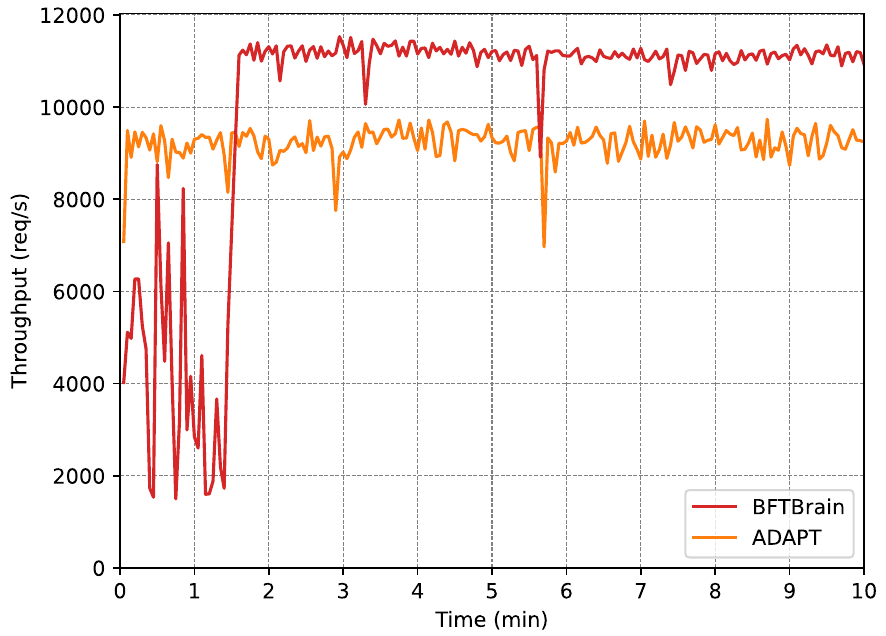}
\vspace{-1em}
\caption{Adaptivity of \sys compared to \adapt under different hardware (WAN).}
\vspace{1em}
\label{fig:wan}
\end{figure}

We compare \sys with state-of-the-art learning based approach \adapt, which is pre-trained with complete data that we collected in this setup when running \sys. Figure~\ref{fig:randomized_cdf} shows the number of committed requests with respect to time. During the first 40 minutes, \adapt performs even slightly better than \sys. This is because \sys incurs a few explorations (i.e., chooses sub-optimal protocols) when starting the system from scratch and the experience buckets are all empty, while \adapt exploits its prior knowledge and thus avoids all such explorations. However, starting from 40 minutes, when the condition varies towards a more Byzantine setup (i.e., with persistent in-dark attacks and leaders that appear randomly slow), \sys gradually outperforms \adapt since the latter does not featurize faults in the system. Thus, although pre-trained with complete data, \adapt consistently picks sub-optimal protocols since its predictive model does not have enough information to tell how the condition changed. As a result, during the entire 2-hour deployment, \sys commits 44\% more requests than \adapt. 

Compared to the 14\% improvement in the ``cycle back'' experiment, \sys improves \adapt by a larger margin in this experiment. The reason is that some input factors in the cycle back experiment are correlated, e.g., a request size near zero is correlated with high proposal slowness. Thus, although \adapt suffers from incomplete features, it indirectly learns the optimal protocol under high proposal slowness using other features, as shown in Figure~\ref{fig:cycleback_cdf} between 90-150 minutes. However, the randomized sampling in this experiment breaks such a correlation.

\subsection{Adaptivity under WAN} \label{sec:appendix_wan}

\begin{figure}[t]
\centering
\includegraphics[width=0.9\linewidth]{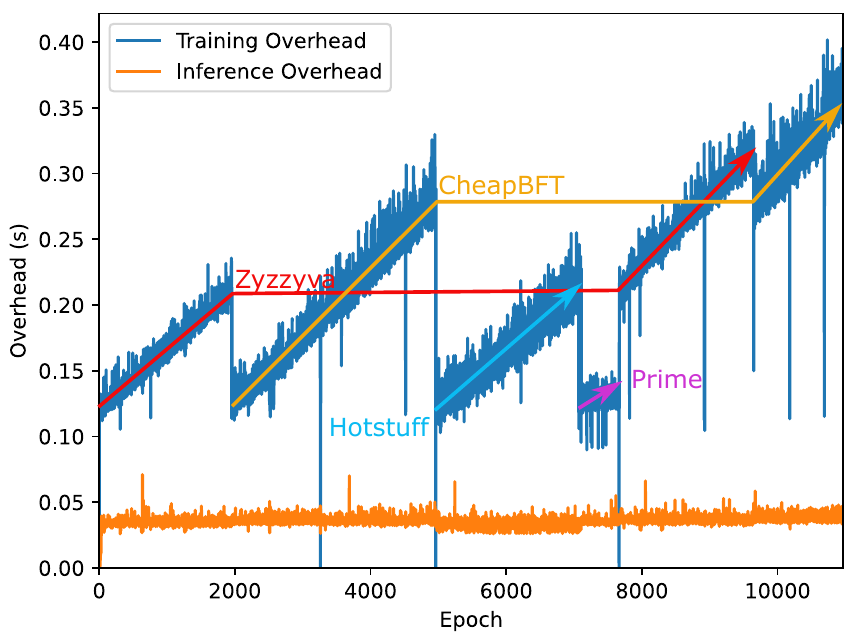}
\vspace{-1em}
\caption{Learning overhead of \sys in each epoch.}
\vspace{1em}
\label{fig:overhead}
\end{figure}

Figure~\ref{fig:wan} plots \sys and \adapt's throughput with respect to time on the ``Row 1 (WAN)'' setup described in Section~\ref{sec:exp_hardware}. Here, \adapt is pre-trained with complete data we collected on ``Row 1 (LAN)'', while \sys is started from scratch. \sys converges to the optimal protocol CheapBFT after a few explorations. On the contrary, \adapt does not perform any explorations and performs relatively well from the very beginning, but is stuck at the sub-optimal decision Zyzzyva. This is because the supervised model \adapt trained is dependent on the hardware. Therefore, \adapt cannot transfer what it learned on a LAN setup to a WAN setup, unless running a prolonged data re-collection process and re-training the supervised model before this new deployment.

\subsection{Overhead of \sys} \label{sec:appendix_overhead}

Figure~\ref{fig:overhead} plots \sys's training and inference overhead in each epoch. Labels indicate the dominant protocol that \sys chooses in each segment of the figure. 

%% file: main.bbl
\begin{thebibliography}{10}

\bibitem{diembft-v4}
The diem team.
\newblock
  https://developers.diem.com/papers/diem-consensus-state-machine-replication-in-the-diem-blockchain/2021-08-17.pdf,
  2021.

\bibitem{thompson_bound}
Shipra Agrawal and Navin Goyal.
\newblock Further optimal regret bounds for thompson sampling.
\newblock In {\em The International Conference on Artificial Intelligence and
  Statistics}, {AISTATS} '13.

\bibitem{allouah2023robust}
Youssef Allouah, Rachid Guerraoui, Nirupam Gupta, Rafael Pinot, and Geovani
  Rizk.
\newblock Robust distributed learning: Tight error bounds and breakdown point
  under data heterogeneity.
\newblock In {\em Conf. on Neural Information Processing Systems (NeurIPS)},
  2023.

\bibitem{amir2011prime}
Yair Amir, Brian Coan, Jonathan Kirsch, and John Lane.
\newblock Prime: Byzantine replication under attack.
\newblock {\em Transactions on Dependable and Secure Computing}, 8(4):564--577,
  2011.

\bibitem{amiri2024bedrock}
Mohammad~Javad Amiri, Chenyuan Wu, Divyakant Agrawal, Amr El~Abbadi, Boon~Thau
  Loo, and Mohammad Sadoghi.
\newblock The bedrock of byzantine fault tolerance: A unified platform for
  $\{$BFT$\}$ protocols analysis, implementation, and experimentation.
\newblock In {\em 21st USENIX Symposium on Networked Systems Design and
  Implementation (NSDI 24)}, pages 371--400, 2024.

\bibitem{androulaki2018hyperledger}
Elli Androulaki, Artem Barger, Vita Bortnikov, Christian Cachin, Konstantinos
  Christidis, Angelo De~Caro, David Enyeart, Christopher Ferris, Gennady
  Laventman, and Yacov Manevich.
\newblock Hyperledger fabric: a distributed operating system for permissioned
  blockchains.
\newblock In {\em European Conf. on Computer Systems (EuroSys)}, pages
  30:1--30:15. ACM, 2018.

\bibitem{attiya2004distributed}
Hagit Attiya and Jennifer Welch.
\newblock {\em Distributed computing: fundamentals, simulations, and advanced
  topics}, volume~19.
\newblock John Wiley \& Sons, 2004.

\bibitem{aublin2015next}
Pierre-Louis Aublin, Rachid Guerraoui, Nikola Kne{\v{z}}evi{\'c}, Vivien
  Qu{\'e}ma, and Marko Vukoli{\'c}.
\newblock The next 700 bft protocols.
\newblock {\em Transactions on Computer Systems (TOCS)}, 32(4):12, 2015.

\bibitem{babay2019deploying}
Amy Babay, John Schultz, Thomas Tantillo, Samuel Beckley, Eamon Jordan, Kevin
  Ruddell, Kevin Jordan, and Yair Amir.
\newblock Deploying intrusion-tolerant scada for the power grid.
\newblock In {\em Int. Conf. on Dependable Systems and Networks (DSN)}, pages
  328--335. IEEE, 2019.

\bibitem{bahsoun2015making}
Jean-Paul Bahsoun, Rachid Guerraoui, and Ali Shoker.
\newblock Making bft protocols really adaptive.
\newblock In {\em Int. Parallel and Distributed Processing Symposium}, pages
  904--913. IEEE, 2015.

\bibitem{baruch2019little}
Gilad Baruch, Moran Baruch, and Yoav Goldberg.
\newblock A little is enough: Circumventing defenses for distributed learning.
\newblock {\em Advances in Neural Information Processing Systems}, 32, 2019.

\bibitem{nonstationary_thompson}
Omar Besbes, Yonatan Gur, and Assaf Zeevi.
\newblock Stochastic multi-armed-bandit problem with non-stationary rewards.
\newblock In {\em Advances in neural information processing systems}, {NIPS}
  '14, pages 199--207.

\bibitem{bessani2014state}
Alysson Bessani, Joao Sousa, and Eduardo~EP Alchieri.
\newblock State machine replication for the masses with bft-smart.
\newblock In {\em Int. Conf. on Dependable Systems and Networks (DSN)}, pages
  355--362. IEEE, 2014.

\bibitem{bootstrapping}
Leo Breiman.
\newblock Bagging predictors.
\newblock In {\em Machine Learning}, Maching Learning '96.

\bibitem{rf}
Leo Breiman.
\newblock Random forests.
\newblock 45(1):5--32.

\bibitem{burrows2006chubby}
Mike Burrows.
\newblock The chubby lock service for loosely-coupled distributed systems.
\newblock In {\em Symposium on Operating Systems Design and Implementation
  (OSDI)}, pages 335--350. USENIX Association, 2006.

\bibitem{cachin2017blockchain}
Christian Cachin and Marko Vukoli{\'c}.
\newblock Blockchain consensus protocols in the wild.
\newblock In {\em Int. Symposium on Distributed Computing (DISC)}, pages 1--16,
  2017.

\bibitem{castro1999practical}
Miguel Castro and Barbara Liskov.
\newblock Practical byzantine fault tolerance.
\newblock In {\em Symposium on Operating Systems Design and Implementation
  (OSDI)}, pages 173--186. USENIX Association, 1999.

\bibitem{castro2002practical}
Miguel Castro and Barbara Liskov.
\newblock Practical byzantine fault tolerance and proactive recovery.
\newblock {\em Transactions on Computer Systems (TOCS)}, 20(4):398--461, 2002.

\bibitem{learned_gc}
Lujing Cen, Ryan Marcus, Hongzi Mao, Justin Gottschlich, Mohammad Alizadeh, and
  Tim Kraska.
\newblock Learned garbage collection.
\newblock In {\em Proceedings of the 4th {ACM} {SIGPLAN} International Workshop
  on Machine Learning and Programming Languages}, {MAPL} @ {PLDI} '20. {ACM}.

\bibitem{thompson_intro}
Olivier Chapelle and Lihong Li.
\newblock An empirical evaluation of thompson sampling.
\newblock In {\em Advances in neural information processing systems},
  {NIPS}'11.

\bibitem{clement2009upright}
Allen Clement, Manos Kapritsos, Sangmin Lee, Yang Wang, Lorenzo Alvisi, Mike
  Dahlin, and Taylor Riche.
\newblock Upright cluster services.
\newblock In {\em Symposium on Operating Systems Principles (SOSP)}, pages
  277--290. ACM, 2009.

\bibitem{clement2009making}
Allen Clement, Edmund~L Wong, Lorenzo Alvisi, Michael Dahlin, and Mirco
  Marchetti.
\newblock Making byzantine fault tolerant systems tolerate byzantine faults.
\newblock In {\em Symposium on Networked Systems Design and Implementation
  (NSDI)}, volume~9, pages 153--168. USENIX Association, 2009.

\bibitem{cohen2022aware}
Shir Cohen, Rati Gelashvili, Lefteris~Kokoris Kogias, Zekun Li, Dahlia Malkhi,
  Alberto Sonnino, and Alexander Spiegelman.
\newblock Be aware of your leaders.
\newblock In {\em Int. Conf. on Financial Cryptography and Data Security},
  pages 279--295. Springer, 2022.

\bibitem{correia2011byzantine}
Miguel Correia, Giuliana~Santos Veronese, Nuno~Ferreira Neves, and Paulo
  Verissimo.
\newblock Byzantine consensus in asynchronous message-passing systems: a
  survey.
\newblock {\em Int. Journal of Critical Computer-Based Systems}, 2(2):141--161,
  2011.

\bibitem{distler2021byzantine}
Tobias Distler.
\newblock Byzantine fault-tolerant state-machine replication from a systems
  perspective.
\newblock {\em ACM Computing Surveys (CSUR)}, 54(1):1--38, 2021.

\bibitem{dobre2013powerstore}
Dan Dobre, Ghassan Karame, Wenting Li, Matthias Majuntke, Neeraj Suri, and
  Marko Vukoli{\'c}.
\newblock Powerstore: Proofs of writing for efficient and robust storage.
\newblock In {\em Conf. on Computer and communications security (CCS)}, pages
  285--298. ACM, 2013.

\bibitem{duplyakin2019design}
Dmitry Duplyakin, Robert Ricci, Aleksander Maricq, Gary Wong, Jonathon Duerig,
  Eric Eide, Leigh Stoller, Mike Hibler, David Johnson, Kirk Webb, et~al.
\newblock The design and operation of $\{$CloudLab$\}$.
\newblock In {\em Annual Technical Conf. (ATC)}, pages 1--14. USENIX
  Association, 2019.

\bibitem{dwork1988consensus}
Cynthia Dwork, Nancy Lynch, and Larry Stockmeyer.
\newblock Consensus in the presence of partial synchrony.
\newblock {\em Journal of the ACM (JACM)}, 35(2):288--323, 1988.

\bibitem{el1985efficient}
Amr El~Abbadi, Dale Skeen, and Flaviu Cristian.
\newblock An efficient, fault-tolerant protocol for replicated data management.
\newblock In {\em SIGACT-SIGMOD symposium on Principles of database systems},
  pages 215--229. ACM, 1985.

\bibitem{el1985availability}
Amr El~Abbadi and Sam Toueg.
\newblock Availability in partitioned replicated databases.
\newblock In {\em SIGACT-SIGMOD symposium on Principles of database systems},
  pages 240--251. ACM, 1986.

\bibitem{farhadkhani2022byzantine}
Sadegh Farhadkhani, Rachid Guerraoui, Nirupam Gupta, Rafael Pinot, and John
  Stephan.
\newblock Byzantine machine learning made easy by resilient averaging of
  momentums.
\newblock In {\em Int. Conf. on Machine Learning (ICML)}, pages 6246--6283.
  PMLR, 2022.

\bibitem{garcia2013intrusion}
Miguel Garcia, Nuno Neves, and Alysson Bessani.
\newblock An intrusion-tolerant firewall design for protecting siem systems.
\newblock In {\em Conf. on Dependable Systems and Networks Workshop (DSN-W)},
  pages 1--7. IEEE, 2013.

\bibitem{garcia2016sieveq}
Miguel Garcia, Nuno Neves, and Alysson Bessani.
\newblock Sieveq: A layered bft protection system for critical services.
\newblock {\em IEEE Transactions on Dependable and Secure Computing},
  15(3):511--525, 2016.

\bibitem{goodson2004efficient}
Garth~R Goodson, Jay~J Wylie, Gregory~R Ganger, and Michael~K Reiter.
\newblock Efficient byzantine-tolerant erasure-coded storage.
\newblock In {\em Int. Conf. on Dependable Systems and Networks (DSN)}, pages
  135--144. IEEE, 2004.

\bibitem{pillars}
Justin Gottschlich, Armando Solar-Lezama, Nesime Tatbul, Michael Carbin, Martin
  Rinard, Regina Barzilay, Saman Amarasinghe, Joshua~B. Tenenbaum, and Tim
  Mattson.
\newblock The three pillars of machine programming.
\newblock In {\em Proceedings of the 2nd {ACM} {SIGPLAN} International Workshop
  on Machine Learning and Programming Languages}, {MAPL} 2018, pages 69--80.
  Association for Computing Machinery.

\bibitem{gramoli2023diablo}
Vincent Gramoli, Rachid Guerraoui, Andrei Lebedev, Chris Natoli, and Gauthier
  Voron.
\newblock Diablo: A benchmark suite for blockchains.
\newblock In {\em European Conf. on Computer Systems (EuroSys)}. ACM, 2023.

\bibitem{guerraoui2010next}
Rachid Guerraoui, Nikola Kne{\v{z}}evi{\'c}, Vivien Qu{\'e}ma, and Marko
  Vukoli{\'c}.
\newblock The next 700 bft protocols.
\newblock In {\em European conf. on Computer systems (EuroSys)}, pages
  363--376. ACM, 2010.

\bibitem{guerraoui2018hidden}
Rachid Guerraoui, S{\'e}bastien Rouault, et~al.
\newblock The hidden vulnerability of distributed learning in byzantium.
\newblock In {\em Int. Conf. on Machine Learning (ICML)}, pages 3521--3530.
  PMLR, 2018.

\bibitem{gueta2019sbft}
Guy~Golan Gueta, Ittai Abraham, Shelly Grossman, Dahlia Malkhi, Benny Pinkas,
  Michael~K Reiter, Dragos-Adrian Seredinschi, Orr Tamir, and Alin Tomescu.
\newblock Sbft: a scalable decentralized trust infrastructure for blockchains.
\newblock In {\em Int. Conf. on Dependable Systems and Networks (DSN)}, pages
  568--580. IEEE/IFIP, 2019.

\bibitem{controlflag}
Niranjan Hasabnis and Justin Gottschlich.
\newblock {ControlFlag}: a self-supervised idiosyncratic pattern detection
  system for software control structures.
\newblock In {\em Proceedings of the 5th {ACM} {SIGPLAN} International
  Symposium on Machine Programming}, {MAPS} '21, pages 32--42. Association for
  Computing Machinery.

\bibitem{hunt2010zookeeper}
Patrick Hunt, Mahadev Konar, Flavio~P Junqueira, and Benjamin Reed.
\newblock $\{$ZooKeeper$\}$: Wait-free coordination for internet-scale systems.
\newblock In {\em Annual Technical Conf. (ATC)}. USENIX Association, 2010.

\bibitem{kapitza2012cheapbft}
R{\"u}diger Kapitza, Johannes Behl, Christian Cachin, Tobias Distler, Simon
  Kuhnle, Seyed~Vahid Mohammadi, Wolfgang Schr{\"o}der-Preikschat, and Klaus
  Stengel.
\newblock Cheapbft: resource-efficient byzantine fault tolerance.
\newblock In {\em European Conf. on Computer Systems (EuroSys)}, pages
  295--308. ACM, 2012.

\bibitem{karimireddy2021learning}
Sai~Praneeth Karimireddy, Lie He, and Martin Jaggi.
\newblock Learning from history for byzantine robust optimization.
\newblock In {\em Int. Conf. on Machine Learning (ICML)}, pages 5311--5319.
  PMLR, 2021.

\bibitem{kirsch2013survivable}
Jonathan Kirsch, Stuart Goose, Yair Amir, Dong Wei, and Paul Skare.
\newblock Survivable scada via intrusion-tolerant replication.
\newblock {\em IEEE Transactions on Smart Grid}, 5(1):60--70, 2013.

\bibitem{kotla2007zyzzyva}
Ramakrishna Kotla, Lorenzo Alvisi, Mike Dahlin, Allen Clement, and Edmund Wong.
\newblock Zyzzyva: speculative byzantine fault tolerance.
\newblock {\em Operating Systems Review (OSR)}, 41(6):45--58, 2007.

\bibitem{ml_index}
Tim Kraska, Alex Beutel, Ed~H. Chi, Jeffrey Dean, and Neoklis Polyzotis.
\newblock The case for learned index structures.
\newblock In {\em Proceedings of the 2018 International Conference on
  Management of Data}, {SIGMOD} '18. {ACM}.

\bibitem{kwon2014tendermint}
Jae Kwon.
\newblock Tendermint: Consensus without mining.
\newblock 2014.

\bibitem{lamport2001paxos}
Leslie Lamport.
\newblock Paxos made simple.
\newblock {\em ACM Sigact News}, 32(4):18--25, 2001.

\bibitem{lamport2009vertical}
Leslie Lamport, Dahlia Malkhi, and Lidong Zhou.
\newblock Vertical paxos and primary-backup replication.
\newblock In {\em symposium on Principles of distributed computing (PODC)},
  pages 312--313. ACM, 2009.

\bibitem{malkhi2019concurrency}
Dahlia Malkhi.
\newblock {\em Concurrency: the works of Leslie Lamport}.
\newblock ACM, 2019.

\bibitem{malkhi2023hotstuff}
Dahlia Malkhi and Kartik Nayak.
\newblock Hotstuff-2: Optimal two-phase responsive bft.
\newblock {\em Cryptology ePrint Archive}, 2023.

\bibitem{decima}
Hongzi Mao, Malte Schwarzkopf, Shaileshh~Bojja Venkatakrishnan, Zili Meng, and
  Mohammad Alizadeh.
\newblock Learning scheduling algorithms for data processing clusters.

\bibitem{bao}
Ryan Marcus, Parimarjan Negi, Hongzi Mao, Nesime Tatbul, Mohammad Alizadeh, and
  Tim Kraska.
\newblock Bao: Making learned query optimization practical.
\newblock In {\em Proceedings of the 2021 International Conference on
  Management of Data}, {SIGMOD} '21.
\newblock Award: 'best paper award'.

\bibitem{thompson_bootstrap}
Ian Osband and Benjamin Van~Roy.
\newblock Bootstrapped thompson sampling and deep exploration.

\bibitem{platania2016choosing}
Marco Platania, Daniel Obenshain, Thomas Tantillo, Yair Amir, and Neeraj Suri.
\newblock On choosing server-or client-side solutions for bft.
\newblock {\em ACM Computing Surveys (CSUR)}, 48(4):1--30, 2016.

\bibitem{singh2008bft}
Atul Singh, Tathagata Das, Petros Maniatis, Peter Druschel, and Timothy Roscoe.
\newblock Bft protocols under fire.
\newblock In {\em Symposium on Networked Systems Design and Implementation
  (NSDI)}, volume~8, pages 189--204. USENIX Association, 2008.

\bibitem{ml_tuning}
Dana Van~Aken, Andrew Pavlo, Geoffrey~J. Gordon, and Bohan Zhang.
\newblock Automatic database management system tuning through large-scale
  machine learning.
\newblock In {\em Proceedings of the 2017 {ACM} International Conference on
  Management of Data}, {SIGMOD} '17, pages 1009--1024. {ACM}.

\bibitem{polyjuice}
Jiachen Wang, Ding Ding, Huan Wang, Conrad Christensen, Zhaoguo Wang, Haibo
  Chen, and Jinyang Li.
\newblock Polyjuice: \{High-Performance\} transactions via learned concurrency
  control.
\newblock {OSDI} '21, pages 198--216.

\bibitem{wang2022bft}
Xin Wang, Sisi Duan, James Clavin, and Haibin Zhang.
\newblock Bft in blockchains: From protocols to use cases.
\newblock {\em ACM Computing Surveys (CSUR)}, 54(10s):1--37, 2022.

\bibitem{wu2022adachain}
Chenyuan Wu, Bhavana Mehta, Mohammad~Javad Amiri, Ryan Marcus, and Boon~Thau
  Loo.
\newblock Ada{C}hain: A learned adaptive blockchain.
\newblock {\em Proc. of the VLDB Endowment}, 16(8):2033–2046, 2023.

\bibitem{balsa}
Zongheng Yang, Wei-Lin Chiang, Sifei Luan, Gautam Mittal, Michael Luo, and Ion
  Stoica.
\newblock Balsa: Learning a query optimizer without expert demonstrations.
\newblock In {\em Proceedings of the 2022 International Conference on
  Management of Data}, {SIGMOD} '22, pages 931--944. Association for Computing
  Machinery.

\bibitem{yin2018byzantine}
Dong Yin, Yudong Chen, Ramchandran Kannan, and Peter Bartlett.
\newblock Byzantine-robust distributed learning: Towards optimal statistical
  rates.
\newblock In {\em Int. Conf. on Machine Learning (ICML)}, pages 5650--5659.
  PMLR, 2018.

\bibitem{yin2019hotstuff}
Maofan Yin, Dahlia Malkhi, Michael~K Reiter, Guy~Golan Gueta, and Ittai
  Abraham.
\newblock Hotstuff: Bft consensus with linearity and responsiveness.
\newblock In {\em Symposium on Principles of Distributed Computing (PODC)},
  pages 347--356. ACM, 2019.

\bibitem{zhang2022reaching}
Gengrui Zhang, Fei Pan, Michael Dang'ana, Yunhao Mao, Shashank Motepalli,
  Shiquan Zhang, and Hans-Arno Jacobsen.
\newblock Reaching consensus in the byzantine empire: A comprehensive review of
  bft consensus algorithms.
\newblock {\em arXiv preprint arXiv:2204.03181}, 2022.

\bibitem{bandit_survey}
Li~Zhou.
\newblock A survey on contextual multi-armed bandits.

\bibitem{lero}
Rong Zhu, Wei Chen, Bolin Ding, Xingguang Chen, Andreas Pfadler, Ziniu Wu, and
  Jingren Zhou.
\newblock Lero: A learning-to-rank query optimizer.
\newblock 16(6):1466--1479.

\end{thebibliography}
